\documentclass[useAMS,usenatbib]{mn2e}
\usepackage{amssymb}
\usepackage[dvips]{graphicx}
\newcommand{\etal}{\hbox{et al.\ }} 

\title[3 to 12 millimetre studies of dense gas towards the western rim of supernova remnant RX\,J1713.7$-$3946]{3 to 12 millimetre studies of dense gas towards the western rim of supernova remnant RX\,J1713.7$-$3946}
\author[Nigel I. Maxted \etal]{Nigel I. Maxted$^{1}$\thanks{E-mail: nigel.maxted@adelaide.edu.au}, Gavin P. Rowell$^{1}$, Bruce R. Dawson$^{1}$,  Michael G. Burton$^{2}$, \newauthor Brent P. Nicholas$^{1}$, Yasuo Fukui$^{3}$, Andrew J. Walsh$^{4}$, Akiko Kawamura$^{3}$, \newauthor Hirotaka Horachi$^{3}$ and Hidetoshi Sano$^{3}$\\ 
$^{1}$School of Chemistry \& Physics, University of Adelaide, Adelaide, 5005,  Australia\\ 
$^{2}$School of Physics, University of New South Wales, Sydney, 2052, Australia\\ 
$^{3}$Department of Astrophysics, Nagoya University, Furocho, Chikusa-ku, Nagoya, Aichi, 464-8602, Japan\\
$^{4}$Centre for Astronomy, School of Engineering and Physical Sciences, James Cook University, Townsville, 4811, Australia}

\begin{document}

\date{Accepted 2012 February 17. Received 2012 February 16; in original form 2011 December 21}
\maketitle


\begin{abstract} 
The young X-ray and gamma-ray-bright supernova remnant RX\,J1713.7$-$3946 (SNR~G347.3$-$0.5) is believed to be associated with molecular cores that lie within regions of the most intense TeV emission. Using the Mopra telescope, four of the densest cores were observed using high-critical density tracers such as CS(J=1-0,J=2-1) and its isotopologue counterparts, NH$_3$(1,1) and (2,2) inversion transitions and N$_2$H$^+$(J=1-0) emission, confirming the presence of dense gas $\gtrsim$10$^4$\,cm$^{-3}$ in the region. The mass estimates for Core\,C range from 40$M_{\odot}$ (from CS(J=1-0)) to 80$M_{\odot}$ (from NH$_3$ and N$_2$H$^+$), an order of magnitude smaller than published mass estimates from CO(J=1-0) observations.

We also modelled the energy-dependent diffusion of cosmic-ray protons accelerated by RX\,J1713.7$-$3946 into Core\,C, approximating the core with average density and magnetic field values. We find that for considerably suppressed diffusion coefficients (factors $\chi=$10$^{-3}$ down to 10$^{-5}$ the galactic average), low energy cosmic-rays can be prevented from entering the inner core region. Such an effect could lead to characteristic spectral behaviour in the GeV to TeV gamma-ray and multi-keV X-ray fluxes across the core. These features may be measurable with future gamma-ray and multi-keV telescopes offering arcminute or better angular resolution, and can be a novel way to understand the level of cosmic-ray acceleration in RX\,J1713.7$-$3946 and the transport properties of cosmic-rays in the dense molecular cores.
\end{abstract}


\section{Introduction} \label{sec:intro}
The origin of galactic cosmic rays (CRs) is mysterious, but popular theories attribute observed fluxes to first order Fermi acceleration in the shocks of supernova remnants (SNRs) (eg. \citet{Blandford:1978}). One such potential CR-accelerator is RX\,J1713.7$-$3946, a young, $\sim$1600 year-old \citep{Wang:1997} SNR that exhibits shell-like properties at both X-ray \citep{Pfeffermann:1996,Koyama:1997,Lazendic:2003,Cassam:2004,Tanaka:2008,Acero:2009} and TeV gamma-ray energies \citep{Aharonian:2006,Aharonian:2007}. 

\begin{figure*}
\centering 
\includegraphics[width=1.00\textwidth]{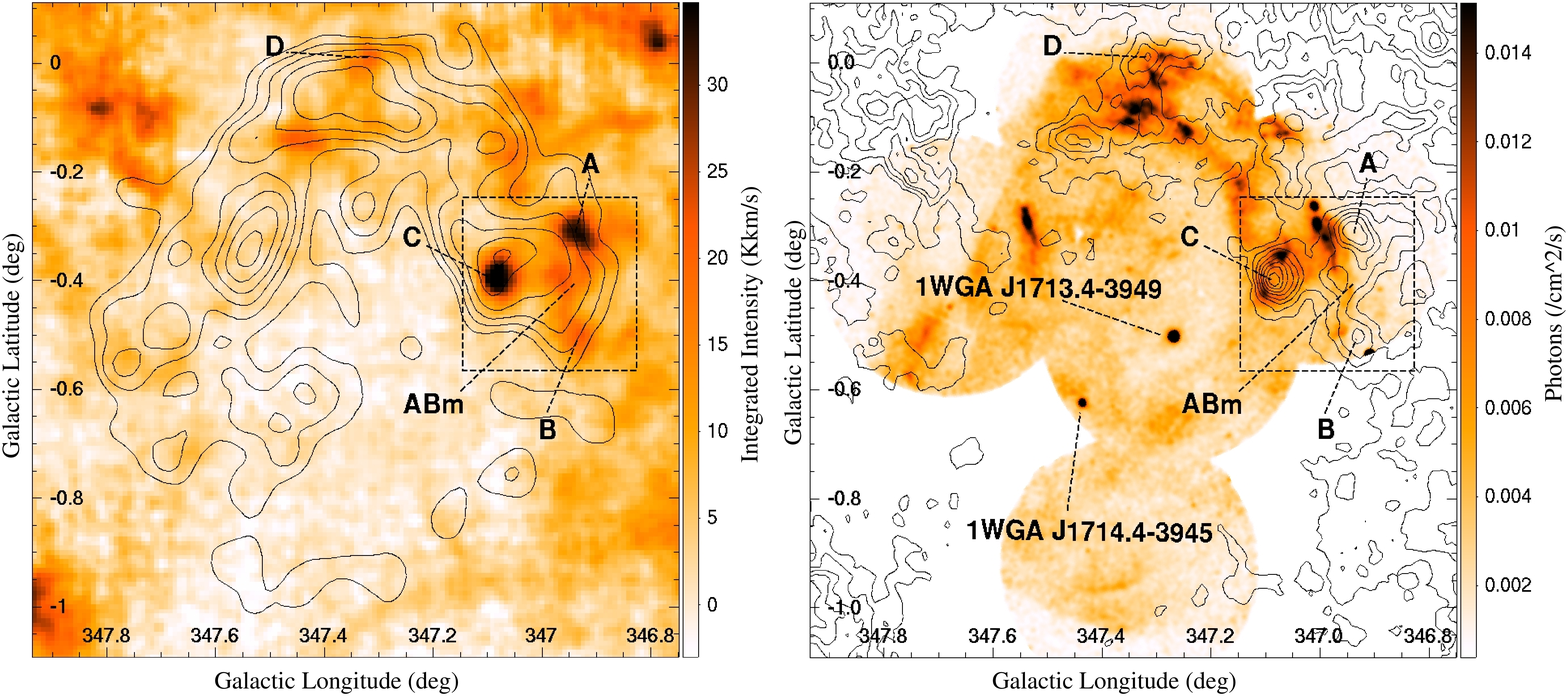} 
\caption{\textbf{a)}(left) Nanten2 CO(J=2-1) integrated-intensity image (v$_{\textrm{\tiny{LSR}}}=-$18~to~0\,km\,s$^{-1}$) \protect\citep{Sano:2010} with HESS $>$1.4\,TeV gamma-ray excess count contours overlaid \protect\citep{Aharonian:2006}. Contour units are excess counts from 20 to 50 in intervals of 5. All core names except ABm were assigned by \protect\citet{Moriguchi:2005}. The dashed 18$^{\prime}\times$18$^{\prime}$ box marks the area scanned in 7\,mm wavelengths by Mopra (Figure~\ref{fig:CSmap}). \textbf{b)}(right) XMM-Newton 0.2-12\,keV X-ray image with Nanten2 CO(J=2-1) (v$_{\textrm{\tiny{LSR}}}$=-18~to~0\,km\,s$^{-1}$) contours. CO(J=2-1) contours span 5\,K\,km\,s$^{-1}$ to 40\,K\,km\,s$^{-1}$ in increments of 5\,K\,km\,s$^{-1}$. The XMM-Newton image has been exposure-corrected and smoothed with a Gaussian of FWHM=30$^{\prime \prime}$. Two compact objects are indicated: 1WGA\,J1713.4$-$3949, which has been suggested to be associated with RX\,J1713.7$-$3946 and 1WGA\,J1714.4$-$3945, which is likely unassociated \protect\citep{Slane:1999,Lazendic:2003,Cassam:2004}. \label{fig:RXJ1}}.
\end{figure*}

A void in molecular gas, most likely created by the RX\,J1713.7$-$3946 progenitor star-winds \citep{Inoue:2012,Fukui:2012}, has previously been observed in CO(J=1-0) and CO(J=2-1) with the Nanten and Nanten2 4\,m telescope \citep{Fukui:2003,Moriguchi:2005,Fukui:2008}, and more recently, HI in the Southern Galactic Plane Survey (SGPS) \citep{McClure:2005,Fukui:2012}.
Three dense clumps, Cores A, B and C (see Figure~\ref{fig:RXJ1}), to the West of this void, and a fourth clump, Core\,D, in the remnant's northern boundary are coincident with regions of high TeV flux. The beam FWHM of $\sim$10$^{\prime}$ for the detection of gamma-rays with HESS would not be able to resolve TeV features of the scale of these molecular cores ($\sim$1$^{\prime}$).

Three of the four stated cores (Cores A, C and D) feature coincident infrared sources commonly associated with star formation \citep{Moriguchi:2005}. Core\,C displays a bipolar nature in CO(J=3-2) and CO(J=4-3) that may be attributed to protostellar activity or outflows \citep{Moriguchi:2005,Sano:2010}.

RX\,J1713.7$-$3946 has a well-defined high energy structure that is shell-like at both keV X-ray and TeV gamma-ray energies \citep{Aharonian:2006}. The TeV emission contours displayed in Figure~\ref{fig:RXJ1}a reveal a clear shell-like structure. The highest fluxes originate from an arc-shaped region in the northern, north-western and western parts of the remnant. The keV XMM-Newton image \citep{Cassam:2004} consists of extended emission regions that form what appears to be two distinct arced shock fronts in the north, north-west and western parts of Figure~\ref{fig:RXJ1}b. 
Two separate shock structures seen at keV energies towards the north-west, believed to be dominated by synchrotron radiation from electrons within the RX\,J1713.7$-$3946 shock, have been considered as an outwards-moving and a reflected, inwards-moving shock \citep{Cassam:2004}.
Various flux enhancements and filaments are evident along these fronts.

The RX\,J1713.7$-$3946 keV emission was first thought to originate from a distance of $\sim$6\,kpc due to an assumed association with molecular gas at line-of-sight velocities, v$_{\textrm{\tiny{LSR}}}\sim-$70\,km\,s$^{-1}$ to $-$90\,km\,s$^{-1}$ \citep{Slane:1999}, but later CO observations and discussion by \citet{Fukui:2003} placed the remnant inside a void in molecular gas at $\sim$1\,kpc distance (v$_{\textrm{\tiny{LSR}}}\sim-$10\,km\,s$^{-1}$). This distance was consistent with an initial estimate based on an X-ray spectral fit and a Sedov model \citep{Pfeffermann:1996}, which was itself consistent with the SNR age \citep{Wang:1997}.

Interestingly, keV X-ray emission peaks (Figure~\ref{fig:RXJ1}b) seem to show some degree of anticorrelation with molecular gas peaks and cores at v$_{\textrm{\tiny{LSR}}}\sim-$10\,km\,s$^{-1}$, as traced by CO (Figure~\ref{fig:RXJ1}b). Two synchrotron intensity peaks on the border of the densest central region of Core\,C are suggestive of compressions associated with the shock \citep{Sano:2010}, while a dip in non-thermal emission from Core\,C's centre may imply that the electron population is unable to diffuse into the core and/or that keV photons cannot escape. A similar scenario may also be applicable to Core\,D, with an X-ray intensity peak also coincident with Core\,D's boundary. Additional gas-keV association may be seen in a line running along an X-ray filament on the eastern border of Core\,A, Point\,ABm (Midpoint of Cores A and B) and Core\,B. These structures possibly suggest gas swept-up by the RX\,J1713.7$-$3946 shock or progenitor wind. \citet{Inoue:2012} modelled a SN shock-cloud interaction and outlined the resultant high energy emission characteristics that were common to both their model and RX\,J1713.7$-$3946. These included parsec-scale keV X-ray correlation with CO emission, subparsec X-ray anti-correlation with CO emission peaks, short-term variability of bright X-ray emission, and stronger TeV gamma-ray emission in regions with stronger CO emission. Despite this strong evidence for an association, a survey by \citet{Cassam:2004} that targeted shock-tracing 1720\,MHz OH maser emission \citep{Wardle:1999} recorded no detections towards this gas complex. This lack of an indication of a molecular chemistry consistent with a shock may point towards a scenario where the gas does not lie within the SNR shell, but given the weight of evidence to the contrary (outlined in this section), the most probable explanations are, as suggested by \citet{Cassam:2004}, that the density is not sufficiently large to cause OH maser emission or that a non-dissociative C-type shock, like that required, does not exist in this region.

Fermi-LAT observations \citep{Abdo:2011} have recently shown that RX\,J1713.7-3946 exhibits a low but hard-spectrum flux of 1-10\,GeV gamma-ray emission, not predicted by early hadronic emission models (ie. pion-producing proton-proton collisions), but consistent with previously-published lepton-dominated gamma-ray models (ie. inverse Compton scattering of TeV electrons) of RX\,J1713.7$-$3946 \citep{Porter:2006,Aharonian:2007,Berezhko:2010,Ellison:2010,Zirakashvili:2010}. However, a hadronic component is possible towards dense cloud clumps \citep{Zirakashvili:2010}, and perhaps more globally if one considers an inhomogeneous ISM \citep{Inoue:2012} into which the SNR shock has expanded. Further support for a global hadronic component may come from consideration of the molecular and atomic ISM gas together \citep{Fukui:2012}. 
Another novel way to discern the leptonic and/or hadronic nature of the gamma-ray emission is to make use of the potential energy-dependent diffusion of CRs into dense cloud clumps or cores 
(e.g. \citet{Gabici:2007}), which may lead to characteristic features in the spectrum and spatial distribution of secondary gamma-rays and synchrotron X-rays from secondary electrons. In \S\,\ref{sec:CRmodel} we therefore looked at a simplistic model of the energy-dependent diffusion of CRs into a dense molecular core similar to Core\,C, in an effort to characterise the penetration depth of CRs in the core and qualitatively discuss the implications for the gamma-ray and secondary X-ray synchrotron emission.

CO observations, such as those already obtained towards RX\,J1713.7$-$3946, are ideal for tracing moderately dense gas ($\sim$10$^{3}$\,cm$^{-3}$), but they do not necessarily probe well dense regions and cores (10$^{5}$\,cm$^{-3}$) that may play an important role in the transport and interactions of high energy particles. Dense gas may also influence SNR shock propagation. To gain a more complete picture, we performed 7\,mm observations of high-critical density molecules to probe the inner regions of cores and search for evidence of ISM disruption by the SNR shocks and star-formation. Towards this goal, we used the Mopra 22\,m single dish radiotelescope in a 7\,mm survey to observe bands containing the dense gas tracer CS(J=1-0), the shock tracer SiO(J=1-0) \citep{Schilke:1997,Martin-Pintado:2000,Gusdorf:2008a,Gusdorf:2008b} and star-formation tracers CH$_3$OH and HC$_3$N \citep{vanDishoeck:1998}. We chose the western region of RX\,J1713.7$-$3946 due to the evidence of shock-interactions with the molecular gas.

Additionally, a 12\,mm observation of NH$_3$ inversion transitions and archival 3mm observations of CS(J=2-1) provided further information towards Core\,C.

\section{Observations} 
An 18$^{\prime}\times$18$^{\prime}$ region (see Figure~\ref{fig:RXJ1}) centred on ($l$,$b$)$=$(346.991, -0.408) that encompasses Cores A, B and C, was mapped by the 22\,m Mopra telescope in the 7\,mm waveband on the nights of the 22nd of April 2009, 23rd of April 2009, 11th of April 2010 and the 21st of April 2010. The Mopra spectrometer, MOPS, was employed and is capable of recording sixteen tunable, 4096-channel (137.5MHz) bands simultaneously when in `zoom' mode, as used here. A list of measured frequency bands, targeted molecular transitions and achieved $T_{\textrm{\tiny{RMS}}}$ levels are shown in Table \ref{Table:bands}. 
\begin{table}
\centering
\small
\caption{The window set-up for the Mopra Spectrometer, MOPS, at 7\,mm. The centre frequency, targeted molecular line, targeted frequency and total mapping noise (T$_{\textrm{\tiny{RMS}}}$) are displayed. \label{Table:bands}} 
\begin{tabular}{|l|l|l|l|}
	\hline
Centre		& Molecular 					& Line		& Map\\
Frequency	& Emission Line					& Frequency & T$_{\textrm{\tiny{RMS}}}$\\
(GHz)		& 								& (GHz)		& (K/ch)\\
	\hline
42.310 		& $^{30}$SiO(J=1-0,v=0) 		& 42.373365	& $\sim$0.07\\
42.500 		& SiO(J=1-0,v=3) 				& 42.519373	& $\sim$0.07\\
42.840 		& SiO(J=1-0,v=2) 				& 42.820582	& $\sim$0.07\\
			& $^{29}$SiO(J=1-0,v=0) 		& 42.879922	& \\
43.125 		& SiO(J=1-0,v=1) 				& 43.122079	& $\sim$0.08\\
43.395 		& SiO(J=1-0,v=0) 				& 43.423864	& $\sim$0.08\\
44.085 		& CH$_3$OH(7(0)-6(1)\,A++) 	& 44.069476	& $\sim$0.08\\
45.125 		& HC$_7$N(J=40-39) 				& 45.119064	& $\sim$0.09\\
45.255 		& HC$_5$N(J=17-16) 				& 45.26475	& $\sim$0.09\\
45.465 		& HC$_3$N(J=5-4,F=5-4) 			& 45.488839	& $\sim$0.09\\
46.225 		& $^{13}$CS(J=1-0) 				& 46.24758	& $\sim$0.09\\
47.945 		& HC$_5$N(J=16-15) 				& 47.927275	& $\sim$0.12\\
48.225 		& C$^{34}$S(J=1-0) 				& 48.206946	& $\sim$0.12\\
48.635 		& OCS(J=4-3) 					& 48.651604	& $\sim$0.13\\
48.975 		& CS(J=1-0) 					& 48.990957	& $\sim$0.12\\
   \hline
\end{tabular}
\end{table}

In addition to mapping, 7\,mm deep ON-OFF switched pointings (pointed observations) were taken on selected regions on the nights of the 23rd of April 2009, 16th of March 2010, 17th of March 2010 and 18th of March 2010, and a 12mm pointed observation was taken on the night of the 10th of April 2010. Archival 3mm Mopra data on Core\,C were utilised in the analysis and were taken on the 30th of April, 2007 in the MOPS wide-band mode (8.3\,GHz band).


The H$_2$O Southern Galactic Plane Survey (HOPS) \citep{Walsh:2008} surveyed the entire Galactic Longitudinal extent of RX~J1713.7$-$3946 down to a Galactic Latitude of -0.5$^{\circ}$, reaching a noise level of T$_{\textrm{\tiny{RMS}}}\sim$0.25\,K/ch. From these data, only a weak NH$_3$(1,1) detection towards Core\,C was found. We took deeper 12mm pointed observations towards this region in response to the HOPS detection, revealing NH$_3$(1,1) and NH$_3$(2,2) emission of a sufficient signal-noise ratio to extract some gas parameters.

The velocity resolutions of the 7 and 12\,mm zoom-mode and 3\,mm wideband-mode data are $\sim$0.2\,km\,s$^{-1}$, $\sim$0.4\,km\,s$^{-1}$ and $\sim$1\,km\,s$^{-1}$ respectively. The beam FWHM of Mopra at 3, 7 and 12\,mm are 36$\pm$3$^{\prime\prime}$, 59.4$\pm$2.4$^{\prime\prime}$ and 123$\pm$18$^{\prime\prime}$, respectively, and the pointing accuracies are $\sim$6$^{\prime\prime}$. The achieved T$_{\textrm{\tiny{RMS}}}$ values for 7\,mm maps and individual 3, 7 and 12\,mm pointings are stated in Tables \ref{Table:bands} and \ref{Table:DeepPointings}.

On-The-Fly (OTF) mapping and pointed observation data were reduced and analysed using the ATNF analysis programs, \textsc{Livedata}, \textsc{Gridzilla}, \textsc{Kvis}, \textsc{Miriad} and \textsc{ASAP}\footnote{See http://www.atnf.csiro.au/computing/software/}. 

\textsc{Livedata} was used to calibrate each OTF-map scan (row/column) using the intermittently measured background as a reference. A linear baseline-subtraction was also applied. \textsc{Gridzilla} then combined corresponding frequency bands of multiple OTF-mapping runs into 16 three-dimensional data cubes, converting frequencies into line-of-sight velocities. Data were weighted by the Mopra system temperature and smoothed in the Galactic $l-b$ plane using a Gaussian of FWHM 1.25$^{\prime}$.

\textsc{Miriad} was used to correct for the efficiency of the instrument \citep{Urquhart:2010} for map data and create line-of-sight velocity-integrated intensity images (moment 0) from data cubes. 

\textsc{ASAP} was used to analyse pointed observation data. Data were time-averaged, weighted by the system temperature and had fitted polynomial baselines subtracted. Like mapping data, deep pointing spectra were corrected for the Mopra efficiency \citep{Ladd:2005,Urquhart:2010}.


\section{Line Detections}

This investigation involved 6 different species of molecule and all detections are displayed in Table \ref{Table:DeepPointings}. Our 7\,mm survey found the dense gas tracers CS, C$^{34}$S and $^{13}$CS in the J=1-0 transition and HC$_3$N(J=5-4) (C$^{34}$S(J=1-0) and HC$_3$N(J=4-5) emission maps are in \S\,\ref{sec:AppOtherMol}). Various detections of CH$_{3}$OH are indicative that these cores have a warm chemistry. Deep 12\,mm observations measured two lines of rotational NH$_3$ emission. In addition to these molecules, archival 3\,mm data revealed detections of the J=2-1 transitions of CS and C$^{34}$S, transitions of SO and N$_2$H$^+$ and further detections of CH$_{3}$OH.

\subsection{CS(J=1-0) Detections}\label{sec:CSdetections}
\begin{figure}
\centering 
\includegraphics[width=0.48\textwidth]{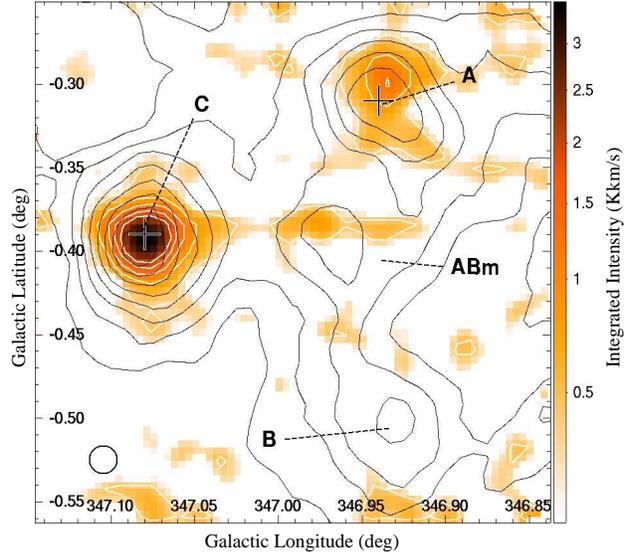}
\caption{Mopra CS(J=1-0) integrated-intensity image (v$_{\textrm{\tiny{LSR}}}$=-13.5 to -9.5\,km\,s$^{-1}$) with Nanten2 CO(J=2-1) (v$_{\textrm{\tiny{LSR}}}$=-18 to 0\,km\,s$^{-1}$) contours (black) and two 8\,$\mu$m point sources indicated by crosses. CO(J=2-1) contours span 5\,K\,km\,s$^{-1}$ to 40\,K\,km\,s$^{-1}$ in increments of 5\,K\,km\,s$^{-1}$. CS(J=1-0) contours (white) are also shown, spanning from 0.4 to 2.0\,K\,km\,s$^{-1}$ in increments of 0.4\,K\,km\,s$^{-1}$. The beam FWHM of Mopra at 49GHz is displayed in the bottom left corner. \label{fig:CSmap}}
\end{figure}

Figure~\ref{fig:CSmap} is a CS(J=1-0) map of the Core\,A-B-C region of RXJ1713.7$-$3946. CS(J=1-0) emission can be seen to be originating from the two most CO(J=2-1)-intense regions, Cores A and C at line-of-sight-velocities of $\sim-$10\,km\,s$^{-1}$ and $\sim-$12\,km\,s$^{-1}$, respectively, consistent with CO emission line-of-sight-velocities. The Core\,C CS(J=1-0) emission is coincident with the Core\,C infrared source, but the Core\,A CS(J=1-0) emission peak is offset from the Core\,A infrared source. This discrepancy of $\sim$1$^{\prime}$, similar to the size of the 7\,mm beam FWHM, is not evident for CO(J=2-1) emission.

\begin{figure}
\centering
\includegraphics[width=0.47\textwidth]{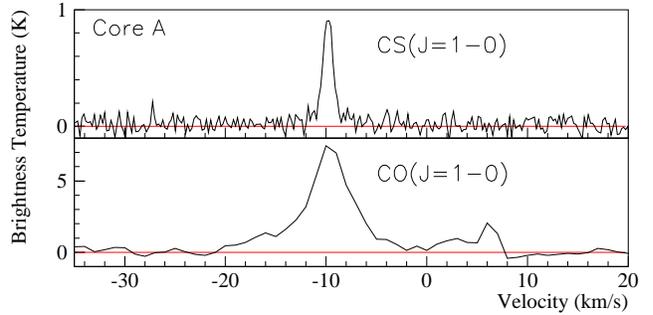}
\caption{\textbf{Core\,A:} CS(J=1-0) spectrum (top) and CO(J=1-0) spectrum (bottom) towards Core\,A. The position of Core\,A is that displayed in Table\,\ref{Table:DeepPointings}. \label{fig:CoreA_CS}}
\end{figure}

\begin{figure}
\centering
\includegraphics[width=0.47\textwidth]{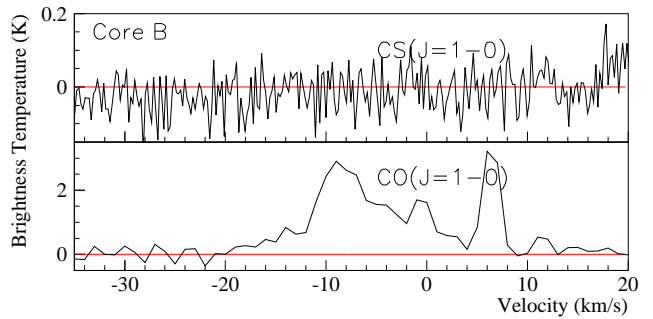}
\caption{\textbf{Core\,B:} CS(J=1-0) spectrum (top) and CO(J=1-0) spectrum (bottom) towards Core\,B. The position of Core\,B is that displayed in Table\,\ref{Table:DeepPointings}. \label{fig:CoreB_CS}}
\end{figure}

\begin{figure}
\centering
\includegraphics[width=0.47\textwidth]{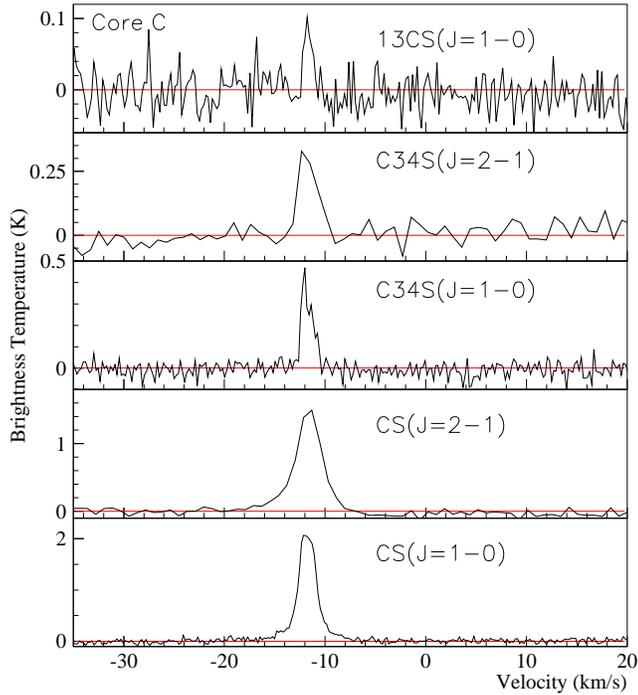}
\caption{\textbf{Core\,C:} CS isotopologue spectra towards Core\,C. The position of Core\,C is that displayed in Table\,\ref{Table:DeepPointings}.\label{fig:CoreC_CS}}
\end{figure}

\begin{figure}
\centering
\includegraphics[width=0.47\textwidth]{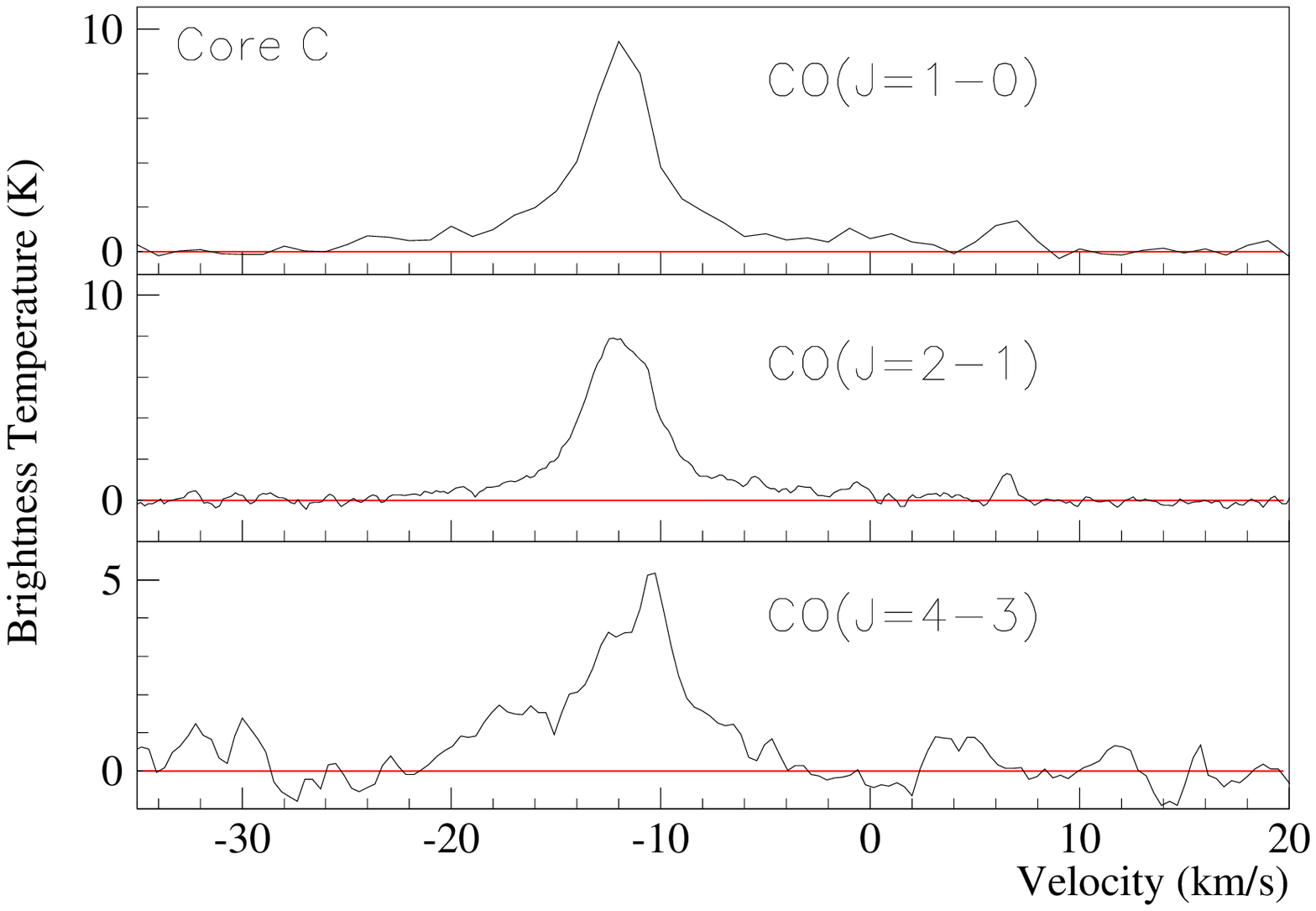}
\caption{\textbf{Core\,C: }Various CO spectra towards Core\,C \citep{Fukui:2003,Moriguchi:2005,Fukui:2008,Sano:2010}. The position of Core\,C is that displayed in Table\,\ref{Table:DeepPointings}. \label{fig:CoreC3_CO}}
\end{figure}

\begin{figure}
\centering
\includegraphics[width=0.47\textwidth]{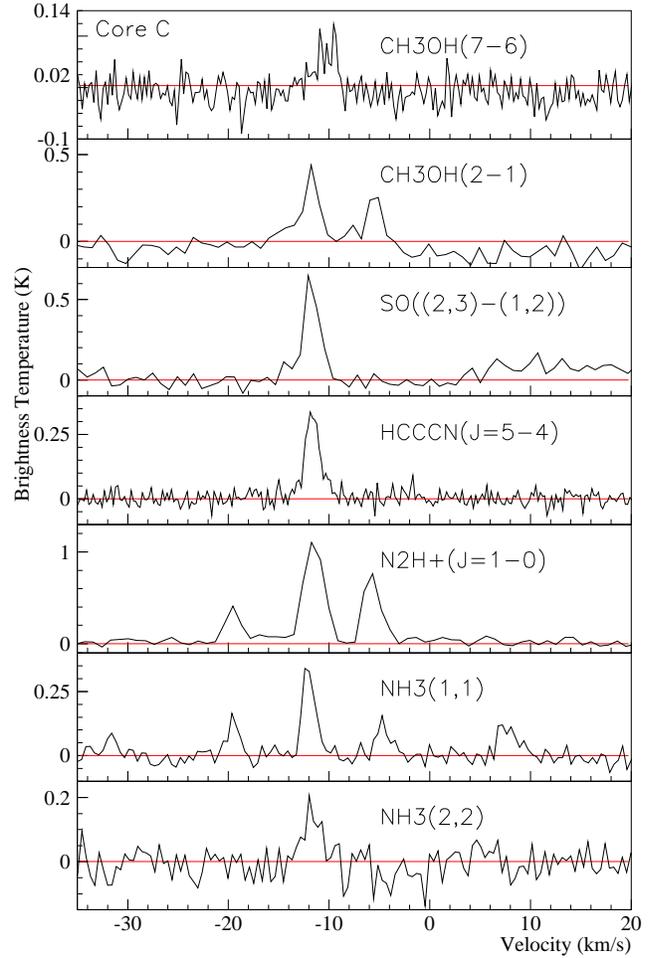}
\caption{\textbf{Core\,C: }Various spectra towards Core\,C. The position of Core\,C is that displayed in Table\,\ref{Table:DeepPointings}. \label{fig:CoreC2_CS}}
\end{figure}

\begin{figure}
\centering
\includegraphics[width=0.47\textwidth]{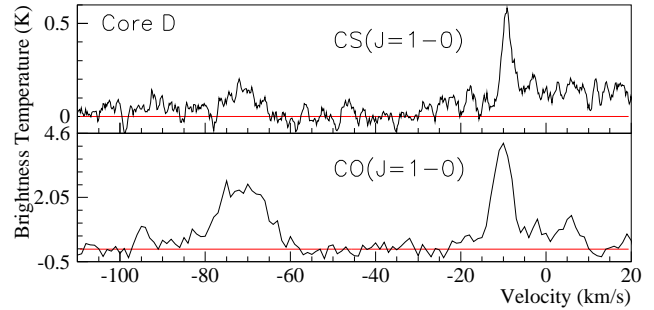}
\caption{\textbf{Core\,D:} CS(J=1-0) spectrum (top) and CO(J=1-0) spectrum (bottom) towards Core\,D. CS(J=1-0) emission has been box-car-smoothed over 6 channels. The position of Core\,D is that displayed in Table\,\ref{Table:DeepPointings}. \label{fig:CoreD_CS}}
\end{figure}

\begin{figure}
\centering
\includegraphics[width=0.47\textwidth]{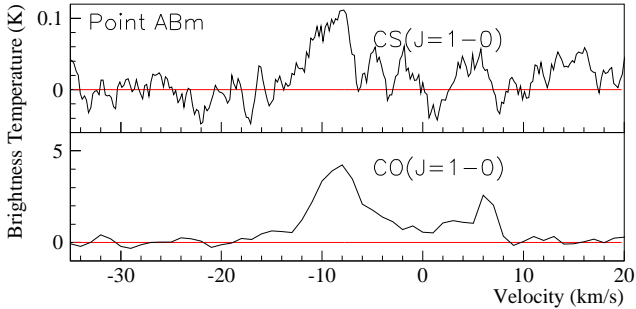}
\caption{\textbf{Point\,ABm:} CS(J=1-0) spectrum (top) and CO(J=1-0) spectrum (bottom) towards Point\,ABm. CS(J=1-0) emission has been box-car-smoothed over 6 channels. The position of Point\,ABm is that displayed in Table\,\ref{Table:DeepPointings}.\label{fig:PointABm_CS}}
\end{figure}

Figures \ref{fig:CoreA_CS}, \ref{fig:CoreB_CS}, \ref{fig:CoreC_CS}, \ref{fig:CoreC2_CS}, \ref{fig:CoreC3_CO}, \ref{fig:CoreD_CS} and \ref{fig:PointABm_CS} show CS(J=1-0) spectra from pointed observations towards Cores A, B, C and D, and Point\,ABm. Corresponding CO(J=1-0) spectra \citep{Moriguchi:2005} are also shown in these figures, but the CO beam FWHM of $\sim$180$^{\prime\prime}$, corresponding to $\sim$9$\times$ the 7\,mm solid angle, means that neighbouring gas is probably included in the beam-average.

\begin{table*}
\small
\caption{Detected molecular transitions from pointed observations. Velocity of peak, v$_{\textrm{\tiny{LSR}}}$, peak intensity, T$_{\textrm{\tiny{Peak}}}$, and FWHM, $\Delta v_{\textrm{\tiny{FWHM}}}$, were found by fitting Gaussians before deconvolving with the MOPS velocity resolution. Displayed values include the beam efficiencies \protect\citep{Urquhart:2010,Ladd:2005}, after a linear baseline subtraction. Statistical uncertainties are shown, whereas systematic uncertainties are $\sim$7\%, $\sim$2.5\% and $\sim$5\% for the 3\,mm, 7\,mm and 12\,mm calibration, respectively \protect\citep{Ladd:2005,Urquhart:2010}. Band noise, T$_{\textrm{\tiny{RMS}}}$, integrated intensity, $\int$T$_{mb}$dv, possible counterparts and 12 and 100\,$\mu$m IRAS flux, $F_{12}$ and $F_{100}$, respectively, (where applicable) are also displayed. \label{Table:DeepPointings}}
\begin{tabular}{|l|l|l|l|l|l|l|l|}
	\hline
Object				& Detected	   		& T$_{\textrm{\tiny{RMS}}}$	&Peak v$_{\textrm{\tiny{LSR}}}$	& T$_{\textrm{\tiny{Peak}}}$	& $\Delta v_{\textrm{\tiny{FWHM}}}$ & $\int$T$_{mb}$dv & Counterparts\\
($l$,$b$)				& Emission Line		& (K/ch)	&(km\,s$^{-1}$)	& (K)           & (km\,s$^{-1}$)	&(Kkms$^{-1}$)& [$F_{12}$/$F_{100}$ (Jy)]\\
	\hline
	\hline
Core\,A 				& CS(J=1-0) 		& 0.06		&-9.82$\pm$0.02	& 0.92$\pm$0.04	& 1.25$\pm$0.06 & 17.0$\pm$0.9 & IRAS 17082$-$3955\\
(346.93$^{\circ}$,-0.30$^{\circ}$)& HC$_3$N(J=5-4,F=5-4)& 0.04 &-9.76$\pm$0.04 & 0.29$\pm$0.02	& 1.07$\pm$0.11	& 4.6$\pm$0.6 & [5.4/138]\\
&SiO(J=1-0) & 0.03&- 				&- 				& - 			&  \\
	\hline
Core\,B				& CS(J=1-0)			& 0.06 	&- 				&- 				& - 			& -\\
(346.93$^{\circ}$,-0.50$^{\circ}$)	& SiO(J=1-0) & 0.03 &- 				&- 				& - 			& -\\
	\hline
					& CS(J=1-0)			& 0.05	&-11.76$\pm$0.01& 2.12$\pm$0.02 &2.08$\pm$0.03 & 65.2$\pm$0.9\\
Core\,C				& CS(J=2-1)			& 0.10	&-11.62$\pm$0.08& 1.49$\pm$0.08	&2.94$\pm$0.21 & 64.8$\pm$2.0 &IRAS 17089$-$3951\\
(347.08$^{\circ}$,-0.40$^{\circ}$)		& C$^{34}$S(J=1-0)	& 0.04 	&-11.83$\pm$0.04& 0.37$\pm$0.02	&1.40$\pm$0.09& 7.7$\pm$0.7 &[4.4/234]\\
					& C$^{34}$S(J=2-1)	& 0.07 	&-11.78$\pm$0.20& 0.34$\pm$0.06	&1.89$\pm$0.49	& 9.5	$\pm$	1.4\\
					& $^{13}$CS(J=1-0)	& 0.03 	&-11.70$\pm$0.10& 0.10$\pm$0.02 &0.88$\pm$0.22 	& 1.3	$\pm$	0.5\\
	 				& HC$_3$N(J=5-4,F=5-4)& 0.02	&-11.65$\pm$0.03& 0.32$\pm$0.01	&1.74$\pm$0.08 & 8.2	$\pm$	0.6\\
					& CH$_3$OH(7(0)-6(1)\,A++)& 0.02 &-10.21$\pm$0.17 & 0.07$\pm$0.01 &2.13$\pm$0.32& 2.2	$\pm$	0.6\\
					& CH$_3$OH(2($-$1)-1($-$1)\,E)& 0.07 & -11.75$\pm$0.17 & 0.33$\pm$0.13 & 0.99$\pm$0.90 & 4.8	$\pm$	1.7\\
					& CH$_3$OH(2(0)-1(0)\,A++) &$^{\prime \prime}$& -11.86$\pm$0.14&0.43$\pm$0.08 	& 1.44$\pm$0.46 & 9.2	$\pm$	1.5\\
					& SO(2,3-1,2)		& 0.07 	&-11.77$\pm$0.09& 0.65$\pm$0.07 &1.74$\pm$0.25 & 16.7	$\pm$	1.4\\
					& N$_2$H$^+$(J=1-0) F$_1$=2-1$^a$& 0.08 &-11.61$\pm$0.03					& 1.14$\pm$0.03	&2.18$\pm$0.08 &36.8	$\pm$	1.1\\
					&~~~~~~~~~~~~~~~~~~~\,F$_1$=0-1& $^{\prime \prime}$ &	$^{\prime \prime}$ 	& 0.40$\pm$0.04	& 1.58$\pm$0.23 &9.4	$\pm$	1.1\\
					&~~~~~~~~~~~~~~~~~~~\,F$_1$=1-1& $^{\prime \prime}$ &	$^{\prime \prime}$ 	& 0.76$\pm$0.03	& 0.87$\pm$0.04 &9.8	$\pm$	0.6\\
					& NH$_3$((1,1)-(1,1)) F$\rightarrow$F$^a$ & 0.03 &-12.04$\pm$0.04 & 0.34$\pm$0.02 &1.46$\pm$0.09 & 7.3	$\pm$	0.7\\
					&~~~~~~~~~~~~~~~~~~~~~~\,F$\rightarrow$F$-$1$^b$ &$^{\prime \prime}$ &$^{\prime \prime}$ & 0.07$\pm$0.02 &$^{\prime \prime}$ &1.5	$\pm$	0.5\\
					& 											  &$^{\prime \prime}$ &$^{\prime \prime}$ & 0.13$\pm$0.02 &$^{\prime \prime}$ &2.8	$\pm$	0.5\\					
					&~~~~~~~~~~~~~~~~~~~~~~\,F$\rightarrow$F+1$^c$  &$^{\prime \prime}$ &$^{\prime \prime}$ & 0.13$\pm$0.02 &$^{\prime \prime}$ &2.8	$\pm$	0.5\\
					& 											  &$^{\prime \prime}$ &$^{\prime \prime}$ & 0.12$\pm$0.02 &$^{\prime \prime}$ & 2.6	$\pm$	0.5\\
					& NH$_3$((2,2)-(2,2)) F$\rightarrow$F$^a$		& $^{\prime \prime}$  &-11.69$\pm$0.10 & 0.16$\pm$0.02  &2.02$\pm$0.26 &4.8	$\pm$	0.8\\
&SiO(J=1-0) & 0.03&- 				&- 				& - 			&  -\\					
	\hline	
Core\,D				& CS(J=1-0)		& 0.10	&-9.06$\pm$0.11	& 0.47$\pm$0.04	&2.50$\pm$0.29	& 17.4	$\pm$	1.3 & IRAS 17078$-$3927\\
(347.30$^{\circ}$,0.00$^{\circ}$)		&				&0.10		&-71.25$\pm$0.46&0.15$\pm$0.04	&3.38$\pm$0.36	& 7.5	$\pm$	1.2 &[2.0/739]\\
&SiO(J=1-0) & 0.06&- 				&- 				& - 			& - \\
	\hline
Point\,ABm			& CS(J=1-0)		& 0.04 &-8.89$\pm$0.43	& 0.09$\pm$0.02	&4~~~$\pm$1~~~&5.3	$\pm$	1.1\\
(346.93$^{\circ}$,-0.38$^{\circ}$)&SiO(J=1-0) & 0.06&- 				&- 				& - 			& - \\
	\hline
\end{tabular}
\\$^a$\textit{Centre line},
$^b$\textit{Outer satellite lines},
$^c$\textit{Inner satellites lines},
\end{table*}

Core\,A exhibits symmetric narrow line CS(J=1-0) emission, of $\Delta v_{\textrm{\tiny{FWHM}}}\sim$1.25\,km\,s$^{-1}$, suggesting that the inner, dense core is not under the influence of an exterior shock. Figure~\ref{fig:RXJ1}b shows that the forward shock is coincident with the outskirts of Core\,A such that the inner CS(J=1-0)-emitting region probably remains unaffected by the RX\,J1713.7$-$3946 shock, consistent with our observations.

Core\,C displays the strongest CS(J=1-0) emission of this survey, implying that this region probably harbours the largest mass dense core. CO(J=4-3) observations by \citet{Sano:2010} suggested that Core\,C contains a bipolar outflow, but we find no obvious asymmetries in the CS(J=1-0) emission line profile or map data. We do, however find a significant difference between the red and blue-shifted sides of CS(J=2-1) emission. For deep switched pointing spectra, fitted gaussians were subtracted from the observed spectra of CS(J=1-0,J=2-1) emission, and the resultant spectra were integrated over velocity ranges chosen, by eye, to represent red and blue-shifted line wings. Integrating CS(J=2-1) emission over 4.1\,kms$^{-1}$-wide bands 3.0\,kms$^{-1}$ either side of the peak emission, highlights a  2-2.5$\sigma$ intensity difference between red and blue-shifted sides, possibly tracing manifestations of the Core\,C bipolar outflow seen by \citet{Sano:2010} in CO emission lines. No significant asymetry ($<$1$\sigma$) is viewed in CS(J=1-0) emission spectra, while in the plane of the sky, there is no significant offset between the red and blue-shifted CS(J=1-0) emission peaks.

In the northern part of the remnant, Core\,D exhibits a CS(J=1-0) emission line with a FWHM of 2.5$\pm$0.3\,km\,s$^{-1}$, slightly larger than that for Cores A and C. This might be expected given that Core\,D harbours an IRAS point source more luminous than that of the other cores associated with RX\,J1713.7$-$3946 \citep{Moriguchi:2005}, but Core\,D also lies on the northern shock front, another possible source of disruption. Towards Core\,D, a second peak at v$_{\textrm{\tiny{LSR}}}=\sim-$70\,km\,s$^{-1}$ ($\sim$6\,kpc) was detected. This Norma-arm gas, initially suggested by \cite{Slane:1999} to be associated with RX\,J1713.7$-$3946, is probably unassociated, given the preference for a 1\,kpc distance (see \S\ref{sec:intro}).

Deep observations on a location at the midpoint of Cores A and B, Point\,ABm, revealed weak and moderately broad CS(J=1-0) emission of FWHM, $\Delta v_{\textrm{\tiny{FWHM}}}$=4$\pm$1, possibly indicating the existence of a passing shock, as there is no evidence to support star-formation towards this location. This scenario is qualitatively supported by the positional coincidence of Point\,ABm with the western shock-front (see \S\ref{sec:intro}), but the low signal to noise ratio means that any conclusions drawn from this are low-confidence. Alternatively, small-scale unresolved clumps (perhaps from star-formation) may be responsible for the detection.

\subsection{Other Molecular species}
No emission from the shock tracer, SiO, was seen in maps or pointed observations. The RX\,J1713.7$-$3946 shock-front likely exceeds 1000\,km\,s$^{-1}$, which may dissociate molecules in the region, but also initiate non-dissociative subsidiary shocks in dense gas, as discussed by \citet{Uchiyama:2010} for older SNRs. Given the existence of dense gas in the region, as indicated by this study, one might indeed expect such secondary shocks if the gas lies within the SNR-shell. In such scenarios, it may be expected that Si is sputtered from dust grains, increasing the gas-phase abundance of SiO in the post-shock region (see for example \citet{Schilke:1997}). As we do not detect SiO towards this region of RX\,J1713.7-3946, we conclude that, if the SN shock is indeed interacting with the surveyed gas as indicated by high energy emissions (see \S\,\ref{sec:intro}), it has not produced SiO at detectable levels.



Detections of the organic molecules HC$_3$N and CH$_3$OH were recorded from Core\,C and HC$_3$N from Core\,A. CH$_3$OH is believed to be the product of grain-surface reactions. Weak emission from this molecule suggests that conditions within the core are capable of evaporating a significant amount of the ice mantles of grains into gas phase.

N$_2$H$^+$ and NH$_3$ were both detected inside Core\,C. NH$_3$ is generally thought to be a tracer of cool, quiescent gas, as its main reaction pathways do not require high temperatures \citep{LeBourlot:1991}. The main reaction pathway of N$_2$H$^+$ is through molecular nitrogen interacting with H$_3^+$ \citep{Hotzel:2004}, so N$_2$H$^+$ can be thought as a tracer of molecular nitrogen. 



\section{Millimetre Emission Line Analyses}
Local Thermodynamic Equilibrium (LTE) analyses using NH$_3$(1,1), NH$_3$(2,2), CS(J=1-0), CS(J=2-1) and N$_2$H$^+$(J=1-0) were applied to estimate gas parameters. For brevity, the mathematics of our molecular emission analyses is not presented here, but in \S\,\ref{sec:AppB}.
Optical depths were found by taking intensity ratios of emission line pairs. Rotational (or excitation in the case of N$_2$H$^+$) temperatures could then be estimated from LTE assumptions allowing total column densities to be calculated. The methods used are only briefly described here, but details are presented in references. After making some assumptions about core structure, masses and densities were estimated.

Gaussian functions were fitted to all emission lines using a $\chi^2$ minimisation method. Lines with visible hyperfine structure (NH$_3$ and N$_2$H$^+$) had each `satellite' line constrained to be a fixed distance (in v$_{\textrm{\tiny{LSR}}}$-space) from the `main' line. In addition to this, for simplicity, all NH$_3$(1,1) emission lines were assumed to have equal FWHM.

\subsection{Optical Depth}\label{sec:OptDepth}
The NH$_3$(1,1) inversion transition has 5 distinctly resolvable peaks and the N$_2$H$^+$(J=1-0) rotational transition has 3 (further splitting is present in both molecular transitions, but remain unresolved due to line overlap). The relative strengths of these lines (increasing in velocity-space) are [0.22, 0.28, 1, 0.28, 0.22] for NH$_3$(1,1) \citep{Ho:1983} and [0.2, 1, 0.6] for N$_2$H$^+$(J=1-0) \citep{Womack:1992}. Here, we refer to the centre lines as `main' lines and all others as `satellite' lines.

To estimate optical depths, $\tau$, of NH$_3$(1,1) and N$_2$H$^+$(J=1-0), any deviations from the aforementioned relative line strengths of main and satellite emission peaks were assumed to be due to optical depth effects, allowing the use of the standard analysis presented in \citet{Barrett:1977}. Note that the optical depths quoted for these two molecular transitions are those of the central emission lines, which in both cases have unresolved structure.

By assuming abundance ratios between CS and rarer isotopologue partners, C$^{34}$S and $^{13}$CS, a similar method can be used to estimate the optical depth of CS emission lines. Values of 22.5 and 75 were assumed for the abundance ratios [CS]/[C$^{34}$S] and [CS]/[$^{13}$CS] respectively. CS emission without detected isotopologue pairs were approximated as being optically thin.



Where multiple optical depth values were calculated (4 for NH$_3$(1,1), 2 for N$_2$H$^+$(J=1-0), 3 for CS(J=1-0,2-1) respectively), an average was taken, weighting by the inverse of the optical depth variance.

\subsection{Temperature and Column Density}
Upper-state column densities, N$_U$, for NH$_3$(1,1), N$_2$H$^+$(J=1-0) and CS(J=1-0,~J=2-1) were calculated via Equation 9 of \citet{Goldsmith:1999}.

\paragraph*{NH$_3$:}
The LTE analysis outlined in \citet{Ho:1983} and summarised in \citet{Ungerechts:1986} was used to find rotational temperature, T$_{rot}$, and total column density, N$_{tot}$, of NH$_3$ from the NH$_3$(1,1) and NH$_3$(2,2) emission lines. Kinetic temperature, T$_{kin}$, was calculated from an approximation presented in \citet{Tafalla:2004}.

\paragraph*{N$_2$H$^+$:}
Given the smaller beam FWHM of the 3mm observations with Mopra (36$^{\prime\prime}$ $<$ 0.2\,pc at 1\,kpc), a filled-beam assumption for the measured N$_2$H$^+$(J=1-0) lines was used to estimate the excitation temperature, T$_{ex}$, in order to find the total column density assuming an approximation of a linear, rigid rotator in LTE \citep{Hotzel:2004}.

\paragraph*{CS:}
Due to the small seperation in energy ($\sim$4.7\,K), where both CS J=1 and J=2 column densities were calculated, a weighted average was taken and assumed to represent the CS(J=1) column density (after a beam-dilution correction, see \S\ref{sec:Beam}). Where possible, an NH$_3$-derived temperature would be assumed, otherwise a temperature of 10\,K was assumed and applied in an LTE approximation to calculate the total CS column density.


\subsection{Beam Dilution}\label{sec:Beam}
A beam dilution factor, $f$, and coupling correction factor, $K$, were applied. Multiplying the column density by the factor:
\begin{equation}
\label{equ:BeamDilution}
fK=\left( 1 - \exp\left[ -\frac{4R^2}{\theta_{mb}^2} \ln{2} \right] \right)^{-1},
\end{equation} corrects for the size of the source compared to the beam FWHM \citep{Urquhart:2010}, where $R$ is the source radius and $\theta_{mb}$ is the main beam FWHM. The main beam FWHM for CS(J=2-1), N$_2$H$^+$(J=1-0), CS(J=1-0) and detected NH$_3$ lines are 36$\pm$3$^{\prime\prime}$, 36$\pm$3$^{\prime\prime}$, 59.4$\pm$2.4$^{\prime\prime}$ and 123$\pm$18$^{\prime\prime}$ respectively \citep{Ladd:2005,Urquhart:2010}. 
$fK$ tends to 1 for beam sizes smaller than the core size.

\subsection{LTE Mass and Density}\label{sec:LTEmass}
LTE masses, $M_{LTE}$, were calculated assuming a spherical, homogeneous core of radius, $R$, composed of molecular hydrogen. Average abundances with respect to H$_2$ for CS, N$_2$H$^+$ and NH$_3$ were assumed to be 1$\times$10$^{-9}$ \citep{Frerking:1980}, 5$\times$10$^{-10}$ \citep{Pirogov:2003} and 2$\times$10$^{-8}$ \citep{Stahler:2005} respectively.

\subsection{Virial Mass}
Under the assumption that turbulent energy balances gravitational energy within a core, a virial mass:
\begin{equation}
\label{equ:Virial}
M_{vir} = k \left(\frac{R}{\textrm{1\,pc}}\right) \left(\frac{\Delta v_{\textrm{\tiny{FWHM}}}}{\textrm{1\,km\,s$^{-1}$}} \right)~~~\textrm{[M$_{\odot}$]},
\end{equation} can be calculated, where $R$ is the radius of the core, $\Delta v_{\textrm{\tiny{FWHM}}}$ is the emission line full-width-half-maximum and $k$ is a coefficient that depends on the density profile of the core. The coefficient $k$ is 444 for a Gaussian density profile \citep{Protheroe:2008}, 210 for constant density and 126 for density, $\rho~\alpha~R^{-2}$ profile \citep{MacLaren:1988}. Recent CO observations suggest that the latter is the best fit \citep{Sano:2010}, so $k$=126 is assumed. A core radius consistent with previous beam dilution assumptions is assumed. For CS emission, virial masses were calculated using the rarer isotopologue C$^{34}$S to minimise the effect of optical depth-broadening.

As the virial mass is not always consistent with LTE mass, the molecular abundances that give an LTE mass equal to that of the virial mass was calculated in each case.

\section{Core Parameters}
The critical density of the J=1-0 transition of CS is $\sim$8$\times$10$^4$\,cm$^{-3}$, implying that the density of regions traced by this emission, in this case, Cores A, C and D, are probably distributed around this value. Cores A, C and D are coincident with infrared sources \citep{Moriguchi:2005}, so the detection of CS(J=1-0) emission from these regions supports the notion that these cores harbour dense warm gas. 

\subsection{Core\,C}
The 3, 7 and 12\,mm observations on Core\,C revealed emission from six different molecular species (see Table~\ref{Table:DeepPointings}). Using three of these molecular species, detailed analyses were performed. Results are displayed in Table~\ref{Table:CoreCparams}, and discussed in the following subsections.

\begin{table*}
\caption{Core\,C parameters derived from three different molecular species assuming a core radius 0.12\,pc. Optical depth, rotational temperature, kinetic temperature, molecular column density, H$_2$ column density, LTE mass, LTE density, virial mass and molecular abundance are denoted by $\tau$, T$_{rot}$, T$_{kin}$, N$_{X}$, N$_{H_2}$, $M_{LTE}$, n$_H$, $M_{vir}$ and $\chi_{vir}$. Statistical errors (rounded to the nearest significant figure) for all variables independent of virial equilibrium assumptions are shown. Systematic errors are not displayed, but are discussed in the text. \label{Table:CoreCparams}}
\begin{tabular}{llllllllll}
\hline
Emission 		&$\tau$ $^a$	& T$_{rot}$ & N$_{X}$ $^{bc}$	&N$_{H_2}$ $^{bcd}$	& $M_{LTE}$ $^{bcd}$& n$_{H_2}$ $^{bcd}$				& $M_{vir}$	$^b$	&$\chi_{vir}$ $^{abc}$\\
Lines  	 		& 				& (K)	&($\times$10$^{13}$\,cm$^{-2}$)&($\times$10$^{22}$\,cm$^{-2}$)	&(M$_{\odot}$)	& ($\times$10$^5$\,cm$^{-3}$) &(M$_{\odot}$)&[X]/[H$_2$]\\
\hline
NH$_3$(1,1)		&1.81$\pm$0.51	&21$\pm$6$^e$ 		& 210$\pm$40	& 11$\pm$2 	& 80$\pm$10	& 3$\pm$0.5	& 30			&5$\times$10$^{-8}$\\
NH$_3$(2,2)		&0.40$\pm$0.08	&			& 					&				& \\
\hline
N$_2$H$^+$(J=1-0)&1.06$\pm$0.23	&$^{bf}$4.5$\pm$0.2	& 5$\pm$1	& 11$\pm$3		& 80$\pm$20 & 3$\pm$1		& 70			&5$\times$10$^{-10}$\\
\hline
CS(J=1-0)		&2.71$\pm$0.29	&		& $^g$5.5$\pm$0.4	& 55$\pm$4 		& 40$\pm$3	& 2$\pm$1		& 30$^h$		&1$\times$10$^{-9}$ $^h$\\
CS(J=2-1)		&		& 			&				&				&					&													& 50$^h$		&7$\times$10$^{-10}$ $^h$\\
\hline
\end{tabular}
$^a$Equal excitation temperature assumption,
$^b$Assuming a source of radius 25$^{\prime \prime}$ ($\sim$0.12\,pc), 
$^c$LTE assumption,
$^d$Assuming abundance ratios stated in \S\ref{sec:LTEmass},
$^e$T$_{kin}=$26$\pm$8
$^f$T$_{ex}$,
$^g$Assuming the temperature calculated from NH$_3$,
$^h$Using the optically thin isotopologue C$^{34}$S.
\end{table*}

\subsubsection{Core\,C Size}\label{sec:DiscLTE}
Using CS(J=1-0) map data, we can estimate the core size by deconvolving the intensity distribution with a Gaussian of FWHM equal to the 7\,mm beam FWHM. We assume that the Core\,C radial density profile is well-approximated by a Gaussian and assume the relation $r^2=s^2+r^{\prime 2}$, where $r$, $s$ and $r^{\prime}$ are the observed radius, beam half-width-half-maximum and deconvolved radius, respectively. By extending the definition of $s$ to be the quadrature sum of both the 7\,mm HWHM (29.7$^{\prime\prime}$) and the Gaussian smoothing HWFM of our maps (37.5 $^{\prime\prime}$), we obtained a dense-core radius of $\sim$25$^{\prime\prime}$. Thus, we assumed a core radius of 0.12\,pc (25$^{\prime\prime}$ at a distance of 1\,kpc) in our calculations, yielding correction factors, $fK$, of 1.3, 1.3, 2.5 and 9.1 for CS(J=2-1), N$_2$H$^+$(J=1-0), CS(J=1-0) and NH$_3$ lines, respectively. There is a $\sim$50\% systematic uncertainty associated with our calculated radius that propagates into LTE mass ($M_{LTE}$), density ($n_{H_2}$), virial mass ($M_{vir}$) and virial abundance ($\chi_{vir}$). We do not account for this systematic in Table\,\ref{Table:CoreCparams}, but discuss it in \S\,\ref{sec:AppScalFact}.

\subsubsection{Mass and Molecular Abundances}\label{sec:mass&abund}
The adoption of molecular abundance ratios is probably the dominant source of systematic uncertainty. It is not uncommon for molecular abundances to vary by an order of magnitude between different Galactic regions, but given the order-of-magnitude agreement between the calculated LTE masses (40-80$M_{\odot}$), the assumed abundances are probably valid. Under the assumption that turbulent energy balances gravitational energy within Core\,C (which may not be valid in a core that is shock-disrupted), virial masses (30 to 70\,M$_{\odot}$) for these three molecular species allowed abundances with respect to molecular hydrogen to be calculated. We saw no significant deviation in CS, N$_2$H$^+$ and NH$_3$ abundance ($\sim$1$\times$, $\sim$1$\times$ and $\sim$2.5$\times$, respectively). 

Systematic uncertainties in beam efficiency, beam-dilution and beam-shape corrections are difficult to account for, but propagating published systematic error estimates leads to greater than 30\% systematic error in N$_2$H$^+$ and CS LTE mass and density estimates, but has only a $\sim$5\% effect on NH$_3$-derived mass and density estimates, since many such effects cancel-out in the analysis.

In our CS analysis of Core\,C, several assumptions were made. The use of an NH$_3$-derived temperature introduces some uncertainty, as NH$_3$ emission may trace a region of different temperature to that of CS emission. \citet{Tafalla:2004} examined the starless cores L1498 and L1517B in detail and found significant variation in CS and NH$_3$ molecular abundances with radius. NH$_3$ (para-NH$_3$) abundance peaked at $\sim$1.4 and 1.7$\times$10$^{-8}$ towards the core centres and dropped to $\sim$5$\times$10$^{-9}$ at $\sim$0.05\,pc, while CS (and C$^{34}$S) abundances decreased towards the centre, with models assuming a central $\sim$0.05\,pc hole, likely due to CS-depletion onto dust grains, providing a better fit to observations than constant-abundance models. It follows that it is feasible that the region emitting in CS(J=1-0) and CS(J=2-1) is different to that emitting in NH$_3$(1,1). The systematic error caused by this effect is difficult to quantify, but given the factor 2 agreement between CS and NH$_3$-derived core mass under a `perfect-correspondence' assumption, it is expected to be small.

The CS J=1 and J=2 states are separated by only 2.35\,K and a comparison of the two emission lines to retrieve a rotational temperature would likely be dominated by uncertain systematics since the 2 measurements were taken with different spectrometers and were separated by 2 years. As we expect the J=1 and J=2 states of CS to be roughly equally-populated, the optical depths and upper-state column densities were averaged to minimise instrumental systematic uncertainties (deemed to be larger than errors introduced by the equal-population assumption).

Of course, all analyses thus far have assumed LTE, which may not be valid, as suggested by CS J=1-0 and J=2-1 intensities. We investigate how the calculated gas parameters may differ from a more accurate non-LTE statistical equilibrium treatment using the freely-available software \textsc{RADEX} \citep{vanderTak:2007}. Through \textsc{RADEX} modeling, we conclude that for a CS column density of 1$\times$10$^{14}$\,cm$^{-2}$, the molecular hydrogen density may not exceed $\sim$6$\times$10$^{4}$\,cm$^{-3}$ at a temperature of 10\,K, and may be smaller for increasing temperature. This solution would have implications for the calculated mass and abundance, being $\sim$75\% smaller and $\sim$3$\times$10$^{-9}$, respectively. We note that in order to correctly simulate the observed $^{13}$CS intensity using the gas parameters outlined above (which were consistent with CS and C$^{34}$S emission), the assumed value of [CS]/[$^{13}$CS] must be adjusted to $\sim$110. Although a limited investigation of CS isotopologue abundance is presented in \S\,\ref{sec:AppIsotop}, we leave a more detailed examination of this for later work.

We note that our measurements seem to trace smaller mass than CO(J=1-0) observations by \citet{Moriguchi:2005}, who calculate a Core\,C mass of $\sim$400\,M$_{\odot}$. This suggests that CS(J=1-0), NH$_3$ and N$_2$H$^+$ are tracing different regions of the molecular cloud to CO, most likely a denser, inner component. The Core\,C radial density profile is discussed further in \S\,\ref{sec:AppRadProf}.



\subsection{Cores A and D}
Table~\ref{tab:CoreADparams} shows gas parameters derived for Cores A and D. We estimate the Core\,A radius to be $\sim$0.3\,pc (60$^{\prime \prime}$ at 1\,kpc) via the method described in \S\ref{sec:DiscLTE}. Core\,A, being more extended than Core\,C, and having an infrared source possibly offset from the CS(1-0) peak emission (see \S\,\ref{sec:CSdetections}), may be composed of two smaller clumps including one emitting in infrared. These imhomogeneities may even be the result of an outflow originating from IRAS\,17082$-$3955, but this scenario remains as speculation. As our maps do not encompass Core\,D, we cannot estimate its radius and assume a radius of 0.12\,pc for Core\,D (see \S\,\ref{sec:AppScalFact} for alternate radius assumptions). 

Given our assumptions, LTE and viral mass solutions for Cores A and D range from 12 to 120\,M$_{\odot}$. We calculate CS abundances a factor 0.2 to 1 of that assumed, and, like in Core\,C, we note that CS(J=1-0) seems to trace less of the total mass than CO(J=1-0). Core\,A and D CO-derived masses are $\sim$700 and $\sim$300\,M$_{\odot}$ respectively \citep{Moriguchi:2005}. Similar to Core\,C, CS likely traces the denser, inner components of Core's A and D.

\begin{table*}
\centering
\caption{CS(J=1-0) parameters for Cores A and D assuming a kinetic temperature of 10\,K. No emission from a rarer isotopologue of CS is detected for these cores, so the maximum optical depth is calculated from the 1\,T$_{\textrm{\tiny{RMS}}}$ noise in the C$^{34}$S band. We display derived values for both optically thin and thick assumptions for parameters that depend on optical depth, giving a range of solutions. Optical depth, molecular column density, H$_2$ column density, LTE mass, LTE density, virial mass and molecular abundance are denoted by $\tau$, N$_{CS}$, N$_{H_2}$, $M_{LTE}$, n$_H$, $M_{vir}$ and $\chi_{vir}$. \label{tab:CoreADparams}}
\begin{tabular}{lllllllllll}
\hline
Core 	&$\tau$ 	& N$_{CS}$ $^{ab}$	& N$_{H_2}$ $^{abc}$	& $M_{LTE}$ $^{abc}$ 	& n$_{H_2}$ $^{abc}$				& $M_{vir}$		&$\chi_{vir}$\\
 	 	&			&($\times$10$^{12}$\,cm$^{-2}$)&($\times$10$^{21}$\,cm$^{-2}$) & (M$_{\odot}$)	& ($\times$10$^3$\,cm$^{-3}$) &(M$_{\odot}$)	&[CS]/[H$_2$]\\
\hline
Core\,A	&0-0.44	& 3-4 		& 3-4 		& 12-15 	& 3-4		& 60	& 2-3$\times$10$^{-10}$\\
Core\,D	&0-4.1	& 40-170 	& 40-170 	& 30-120	& 100-500 	& 100	& 3-10$\times$10$^{-10}$\\
\hline
\end{tabular}
\\$^a$\textit{Assumed rotational temperature of 10\,K,}
$^b$\textit{Corrected for beam dilution assuming a source of radius of 60$^{\prime \prime}$ ($\sim$0.30\,pc) for Core\,A and 25$^{\prime \prime}$ ($\sim$0.12\,pc) for Core\,D,}
$^c$\textit{Assuming abundance ratios stated in \S\ref{sec:LTEmass}}
\end{table*}

\section{High Energy Particle Propagation into dense Cores}\label{sec:CRmodel}
In the context of understanding the nature of the TeV gamma-ray emission from SNRs, of considerable interest is the energy-dependent propagation of high energy protons (CRs) and electrons into dense cores, which may lead to characteristic gamma-ray spectra from GeV to TeV energies. This effect could offer a new way to identify the parent particles (hadronic vs. leptonic) responsible for gamma-ray emission, provided such cores can be spatially resolved in gamma-rays. \citet{Gabici:2007} investigated the propagation of CR protons into molecular clouds and found that in some models, suppressed diffusion of low energy CRs resulted in a gamma-ray spectrum towards the cloud centre that is harder than that towards the cloud edges. This effect became more pronounced for more centrally-condensed gas distributions. Similar phenomena may in fact be occurring towards the dense cores associated with RX\,J1713.7$-$3946. 


We therefore investigated CR diffusion into a molecular core located adjacent to a CR accelerator. Our model is a simplistic approach that only aims to show in principle the effects that dense gas could have on the spatial distribution of CR protons of different energies, which will have knock-on effects on the gamma-ray and X-ray spectra from various parts of the core. 
The model configuration is based on the Core\,C molecular core, and its apparent location with respect to RX\,J1713.7$-$3946 as an assumed CR accelerator. CR transport in this region is expected to be in the diffusion regime, since CR (proton) gyroradii at TeV energies are expected to be $<$10$^{-4}$\,pc for typical magnetic field strengths of 10\,$\mu$G or more.

Our model core is placed 5\,pc from the centre of  RX\,J1713.7$-$3946, which has an outer shock radius growing with time to eventually reach the core, as in the situation assumed for Core\,C. We then calculated the propagation distance (via diffusion) of CR protons of energy, $E_P$, from the outer shock of the SNR, towards the centre of this molecular core, applying time limits based on assumptions concerning the age of RX\,J1713.7$-$3946, and the escape time of CRs from the SNR outer shock. 

\paragraph*{The Molecular Core:}
We modelled a core with radius set equal to 0.62\,pc and a density of 300\,cm$^{-3}$, consistent with volume-averaged density estimates from CO(J=2-1) measurements by \citet{Sano:2010} for Core\,C.
The diffusion of protons is dependent on the strength of magnetic turbulence within the region. These magnetic fields are generally assumed to be frozen-in to the gas, such that denser gas implies larger magnetic field strengths, $B(n_{H_2})$. If we assume the relation stated in \citet{Crutcher:1999}:
\begin{equation}
\label{equ:Crutcher}
B(n_{H_2})\sim 100 \sqrt{ \frac{n_{H_2}}{10^4\,\textrm{\small{cm}}^{-3}} }~~~\textrm{\small{[$\mu$G]}}
\end{equation} the average magnetic field strength inside the core is $B\sim$17\,$\mu$G.

\subsection{Cosmic-Ray Proton Transport}
In the hadronic scenario for gamma ray production, CR protons diffuse from acceleration sites and collide with matter to create TeV emission. The rate of diffusion of CRs is governed by magnetic turbulence, the characteristics of which are somewhat uncertain.

\begin{figure}
\includegraphics[width=0.47\textwidth]{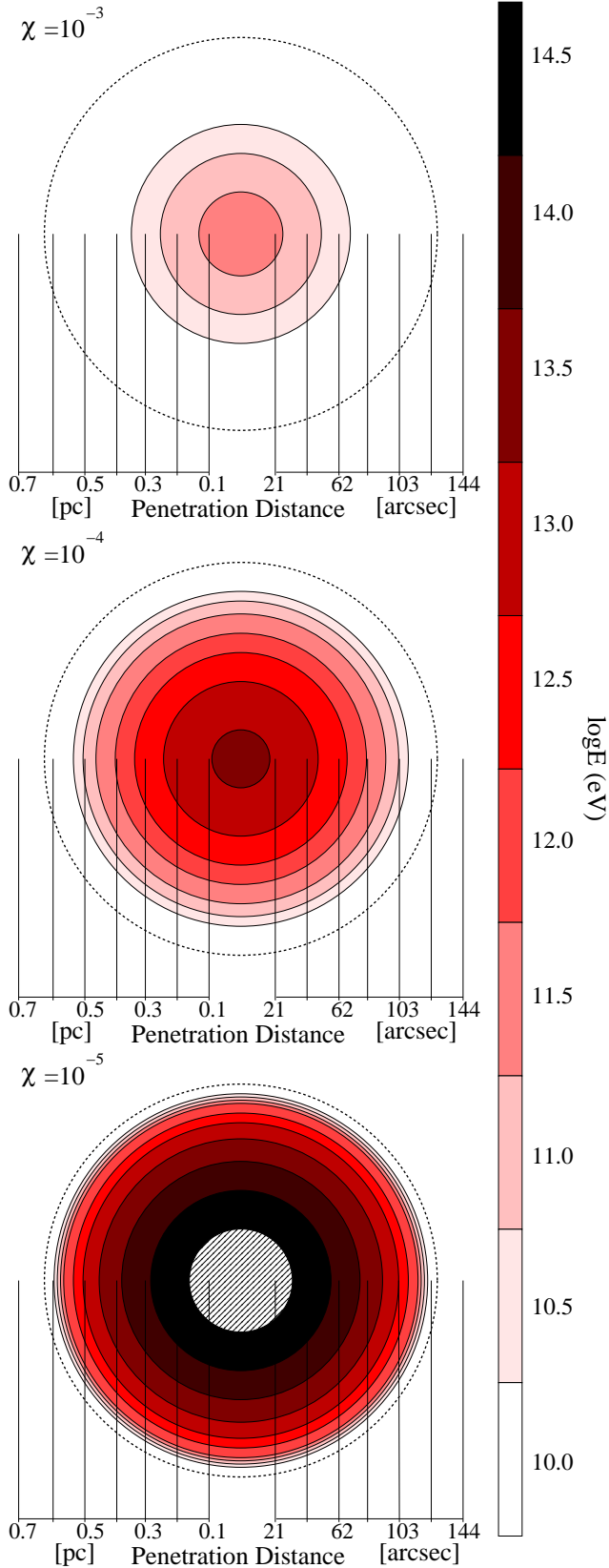}
\caption{The logarithm of the minimum energy CR proton able to penetrate into different radii of our simulated core. The results, presented as 2D slices, from 3 different diffusion suppression coefficients, $\chi$ are displayed. Note that the arcsec scale for penetration distance assumes a distance of 1\,kpc. Hatching indicates a lack of CR inhabitance of CRs. \label{fig:CRpenetrationHom}}
\end{figure}

If we assume that RX\,J1713.7$-$3946 is $\sim$1600 years old \citep{Wang:1997}, we have a limit on the amount of time CR protons have had to be accelerated in the SNR shock and diffuse into the core. Our goal was to examine how far CR protons of different energies penetrated into the core. In Figure\,\ref{fig:CRpenetrationHom}, we have visualised the results (a slice view of the core) of the following equation, which gives the average radial penetration distance of a CR proton within our modelled molecular core. 
\begin{equation}
\label{equ:Penetrate}
R=0.62-\sqrt{6D(E_P,B)[1600-t_0]}~~~\textrm{\small{[pc]}}
\end{equation} where $D(E_P,B)$ is the diffusion coefficient that depends on CR energy, $E_P$, and magnetic field strength, $B$. Equation\,\ref{equ:Penetrate} returns the average distance a CR of energy, $E_P$, will penetrate into the modelled molecular core given 1600 years.

The start time, $t_0$, was set equal to the minimum of the average time taken for the CR proton to escape the SNR shock, $t_{esc}$, and the time taken for the SNR shock to reach the boundary of the core $t_{core}$.
\begin{equation}
\label{equ:tstart}
t_0=\min(t_{esc},t_{core})~~~\textrm{\small{[yr]}}
\end{equation}
The value for $t_{core}\sim$600\,years was based on the Sedov solution for a core-collapse supernova explosion into a wind-driven bubble, where the shock radius evolution is proportional to $t^{-7/8}$ \citep{Ptuskin:2005}. The time-dependent shock radius and energy-dependent escape time mean that some CRs will escape the shock before it contacts the core. After the shock has passed the core ($\sim$600\,years), it is assumed that CRs of all energies can reach the core boundary. In general however for the moment we ignore this CR diffusion time from the shock to the core, which is at most $\sim$100\,yr for the highest energy CRs since it is assumed to take place in a relatively low magnetic ($B\sim$few\,$\mu G$) and density ($n\lesssim 1$\,cm$^{-3}$) environment.  

\paragraph*{Escape Time:}
With the assumption that within a SNR the highest energy particles ($\gg$1\,TeV) are accelerated in the transition from the free-expansion phase to the Sedov phase \citep{Ptuskin:2005} within a short timescale \citep{Uchiyama:2007}, we used the following parameterisation for the CR proton escape time, \citep{Gabici:2009}:
\begin{equation}
\label{equ:CRrelease} 
t _{esc}(E_P) = t_{Sedov}\left( \frac{E_P}{E_{P,max}}\right)^{-1 / \delta}~~~\textrm{\small{[yr]}}
\end{equation} where $E_{P,max}$ is the maximum possible CR proton energy, set equal to 5$\times$10$^{14}$\,eV following \citet{Caprioli:2009} and \citet{Casanova:2010}, $t_{Sedov}$ is the time for the onset of the Sedov phase, set equal to 100\,years, and $\delta$ is a phenomonological parameter that describes the energy-dependent release of CRs. $\delta$ is set equal to 2.48 assuming that $\sim$10$^{15}$\,eV CRs are accelerated near the start of the Sedov phase reducing to $\sim$10$^{9}$\,eV near the end (see Gabici \etal 2009).

\paragraph*{Diffusion:}
\citet{Gabici:2007} considered the timescales involved in different CR transport processes within a dense 20\,pc-diameter molecular cloud. They demonstrated that the dynamical (free-fall) and advective (turbulent) timescales involved were likely an order of magnitude larger than the time-scale for inelastic proton-proton interactions and a varying amount larger than diffusion time-scales. For our model core, we calculated dynamical and advective timescales of $>$10$^4$\,years, so we too ignore these mechanisms and only consider proton-proton interactions (see later) and diffusion, using a diffusion coefficient parameterised as:
\begin{equation}
\label{equ:Diffcoeff}
D(E_P,B(r))=\chi D_0 \left( \frac{E_P/\textrm{\small{GeV}}}{B/3\,\textrm{\small{$\mu$G}}}\right)^{0.5}~~~\textrm{\small{[cm$^2$\,s$^{-1}$]}},
\end{equation} where $D_0$ is the galactic diffusion coefficient, assumed to be 3$\times$10$^{27}$\,cm$^2$\,s$^{-1}$ to fit CR observations \citep{Berezinskii:1990}, and $\chi$ is the diffusion suppression coefficient (assumed to be $<$1 inside the core, 1 outside), a parameter invoked to account for possible deviations of the average galactic diffusion coefficient inside molecular clouds \citep{Berezinskii:1990,Gabici:2007} which is largely unknown.

\paragraph*{Interactions:}
For energies above $\sim$300\,MeV proton-proton (p-p) interactions dominate over ionisation interactions \citep{Gabici:2007}, so we consider the rate of hadronic CR interactions with molecular gas protons and ignore ionisation losses. Assuming an inelasticity of 0.45 (2 interactions give $\sim$79\% energy loss) and a proton-proton cross section of 40\,mb, appropriate for TeV energies, the lifetime of this proton-proton interaction process is:
\begin{equation}
\label{equ:protonlife} 
\tau _{pp} = 6 \times 10^5 \left( \frac{n_H(r)}{100\,\textrm{\small{cm}}^{-3}} \right)^{-1}~~~\textrm{\small{[yr]}}, \label{eq:taupp}
\end{equation} Thus, CRs in the core have a large proton-proton interaction lifetime of $\sim$10$^{5}$\,yr, so we do not model this process. 


\paragraph*{Diffusion Suppression and Results:}
\label{sec:Diffusion}
Diffusion-suppression coefficients $\chi=$ 0.1, 0.01, 10$^{-3}$, 10$^{-4}$ and 10$^{-5}$ were trialled for proton energies 10$^{10}$, 10$^{10.5}$, up to 10$^{14}$\,eV. This wide range of $\chi$ values were chosen to investigate the level of CR penetration with results from the smaller $\chi$ values displayed in Figure\,\ref{fig:CRpenetrationHom}.

For $\chi$=0.1 and 0.01, all CR energies were able (on average) to reach the inner core within the age of the system. For $\chi$=10$^{-3}$, some energy-dependent penetration could be seen, as CRs of energies 10$^{10}$ and 10$^{10.5}$\,eV did not reach the inner core. Energy separation became more prominent with smaller values of $\chi$, as illustrated in Figure\,\ref{fig:CRpenetrationHom}. For $\chi$=10$^{-4}$ and $\chi$=10$^{-5}$, CRs of energy 10$^{12.5}$ to $>10^{14}$ are now excluded from the central regions of the core. This effect would have implications for the spectrum of gamma-ray emission, with the TeV gamma-ray spectrum becoming progressively harder towards the core's centre.


It is unclear whether such extreme cases of diffusion-suppression down to $\chi$=10$^{-5}$ required to produce these energy-dependent penetration effects on such small scales are plausible. Radio continuum measurements by \citet{Protheroe:2008} when compared to the GeV gamma-ray emission have suggested upper limits for $\chi<$0.02 to $<$0.1 (dependent of the assumed magnetic field) inside the Sgr\,B2 molecular cloud (see also \citet{Jones:2011}). Suppressed CR diffusion ($\chi \sim 0.01$ to 0.1) is also indicated when explaining the GeV to TeV gamma-ray spectral variation seen in the vicinity of several evolved SNRs (e.g. \citet{Gabici:2010}). 
Moreover it has been recognised that suppressed diffusion could be expected in turbulent magnetic fields likely to be found in and around shock-disrupted regions (e.g. \citet{Ormes:1988}). Overall, we regard these uncertainties in particle transport properties as motivation for investigating the effects of a wide range of diffusion suppression coefficients, as applied to our model core which may also be similarly shocked and turbulent.

From a more theoretical standpoint, the level of CR diffusion suppression that may be expected is still somewhat unclear. \citet{Yan:2012} recently investigated the effect of CR-induced streaming instabilities, distortions of magnetic field lines caused by the flow of CRs, on CR acceleration within SNRs. The authors expressed interest in solving for a diffusion coefficient in the region surrounding SNRs by incorporating both streaming instabilities and background turbulence. Their future work will help to constrain the level of CR diffusion-suppression that can be plausibly expected inside gas associated with SNRs.


We also note that an additional CR component from the Galactic diffuse background `sea' of CRs will be incident on the core. Given this is an ever-present component, the penetration of these diffuse CRs will not be limited by SNR age, but by the age of the core, and possibly by the energy loss timescale $\tau _{pp}$ due to proton-proton collisions (Eq.\ref{eq:taupp}). For our model core with average density $n=$300\,cm$^{-3}$, $\tau _{pp}$ will not be the dominant loss mechanism, but will become more significant in the central region ($\tau _{pp}\sim$600\,yr for $n=$10$^5$\,cm$^{-3}$). The dynamical age of Core\,C is estimated at $\sim$10$^5$\,yr \citep{Moriguchi:2005} which is similar to the energy loss timescale for our averaged-density model core. In most of the $\chi$ values we consider, these timescales are significantly larger than the time taken for CRs to diffuse into the core centre. For an example extreme case, it takes $\sim$5$\times$10$^5$\,yr for a 10$^{10}$\,eV CR to reach the core centre for $\chi=$10$^{-5}$, so diffuse CRs will likely penetrate the core. However, the energy density of diffuse CRs, assumed to be approximately similar to the Earth-like level of CRs, would be considerably smaller (a factor $\sim$10$^{-4}$) than the local CR component due to an adjacent SNR (e.g. Aharonian \& Atoyan 1996), so we can therefore neglect it here. 


Our core model reflects only average values of density and magnetic field as so will have limitations in comparison to a core where these quantities vary with radius. Compared to the case for a varying diffusion coefficient resulting from a radially dependent density (see Equations \ref{equ:Crutcher} and \ref{equ:Diffcoeff}), our homogenous core model may under-estimate the CR penetration depth in the outer region of the core, and over-estimate the penetration depth towards the inner region of the core. In future work we will consider CR propagation into a core with a radially varying diffusion coefficient based on the measured power-law radial density profile from CO measurements $n(r) \sim 10^5 / (1+(r/0.1{\rm pc})^{2.2})$\,cm$^{-3}$ \citep{Sano:2010}. This model should be three-dimensional and predict a line-of-sight view of gamma/X-ray emission morphology towards the core.

Another limitation concerns the implications of the extremely low values of diffusion suppression we have used down to $\chi=10^{-5}$. For such cases, which represent highly
turbulent magnetic fields, the possibility of second order Fermi reacceleration of the CRs might apply as discussed by \citet{Dogiel:1990}.
\citet{Uchiyama:2010} have in fact considered this effect in accounting for the GeV to TeV gamma-ray spectra towards several evolved SNRs 
(W51C, W44 and IC\,443) by considering the reacceleration of diffuse CRs inside clouds disturbed by a supernova shock. Such effects may 
occur in molecular cores associated with RX\,J1713.7$-$3946, particularly those contacted by the SNR shock  but at this stage we leave this
as an open question. Finally, there may be a regular (or ordered) magnetic field component, which if present may either assist or 
hamper the penetration of CRs depending on the strength and orientation of this field component. Such modeling is beyond the scope of this investigation, but should be considered in more detailed models.

\subsection{Gamma-rays from Core\,C}
\label{sec:Gamma}
We can expect broad-band GeV to TeV gamma-ray emission from CRs penetrating the core as they interact with core protons. The level of the gamma-ray emission will reflect the energy and spatial distribution of CRs within the core coupled to the mass of protons they interact with (see for e.g. \citet{Gabici:2007}). Variations in the gamma-ray spectrum will reflect the energy dependent penetration depth of CRs, and would be a feature only resolvable in gamma-rays with an angular resolution of 1\,arcmin or better according to the size of the cores. The $\sim$6\,arcmin angular resolution offered by H.E.S.S. is at present not sufficient to probe for such spectral variations. 

For a wide range of diffusion suppression factors $\chi=10^{-5}$ to $10^{-3}$, our simple model of Core\,C suggests that lower energy CRs can be prevented from reaching its centre. For the more severe values of $\chi=10^{-4}$ and $10^{-5}$, the CR penetration to the core centre can be limited to CRs of TeV or greater energies, which could lead to considerable differences in the gamma-ray spectra seen towards the centre and edge of the core.

The overall gamma-ray flux expected from dense cores towards RX\,J1713.7$-$3946 has been discussed by \citet{Zirakashvili:2010} (see their Fig.14). They suggested an enhanced TeV hadronic component towards molecular cloud cores (with combined mass $\sim$1000\,M$_\odot$) may be found at a level of $\sim$6$\times$10$^{-12}$\,erg\,cm$^{-2}$s$^{-1}$ for energies above $\sim$10\,TeV, which would result from CRs of energy above $\sim$100\,TeV entering the core, which could be the case for most levels of diffusion suppression. For these gamma-ray energies the dense-core hadronic component may reach or even exceed the SNR-wide leptonic inverse-Compton component. The $>$10\,TeV gamma-ray flux from the central regions of Core\,C, which is 40 to 80\,M$_\odot$ based on our measurements, may therefore reach a few$\times$10$^{-13}$\,erg\,cm$^{-2}$s$^{-1}$, possibly making it detectable by future more sensitive gamma-ray telescopes such as the Cherenkov Telescope Array (CTA) \citep{CTA:2010}. With its expected 1\,arcmin or better angular resolution, CTA could also potentially probe for energy dependent penetration of the parent CRs, by examining the flux and spectra from the inner to outer regions of the core.

\subsection{Electron Transport and X-ray Emission Towards Core\,C}
As discussed in \S\ref{sec:intro}, two synchrotron X-ray emission peaks can be observed towards the boundary of Core\,C (Figure \ref{fig:RXJ1}b). These peaks are perhaps due to increased shock interaction with the dense gas (see \citet{Sano:2010,Inoue:2012}) giving rise to a region of enhanced magnetic field and/or electron acceleration. The observed synchrotron X-ray emission can be produced either by electrons accelerated directly by the SNR shocks outside of Core\,C or, by the secondary electrons from CR proton interactions with ambient gas. In the latter case, the X-ray emission spectrum could be influenced by the energy dependent penetration of CRs into the core and for strong CR diffusion suppression ($\chi\leq10^{-4}$), one might expect an X-ray spectral hardening towards the core centre, in the same way as for gamma-rays discussed in \S\ref{sec:Gamma}.

The dense gas and high column density from Core\,C will also suppress the X-ray flux for energies less than $\sim$5\,keV via photoelectric absorption. Thus some X-ray emission may actually originate from within or behind Core\,C. The X-ray flux decrease seen in going from the boundary of Core\,C to its centre is $\sim$95\%. Assuming the cross section for photoelectric absorption by hydrogen, $\sigma _{\nu}\sim$10$^{-23}$\,cm$^{2}$ (at $\sim$1\,keV), a flux decrease of 95\% corresponds to a traversed column density of N$_H\sim$3$\times$10$^{23}$\,cm$^{-2}$, similar to the column density towards Core\,C from our CS, N$_2$H$^+$ and NH$_3$ observations which pertain to the core centre (See Table\,\ref{Table:CoreCparams}). Therefore it is possible that keV X-ray emission, either from secondary electrons produced deep within Core\,C from penetrating CRs or from external electrons directly accelerated outside and then entering Core\,C, are absorbed by gas within the core. 

A potential way to discriminate between these two scenarios would be to have arcmin angular resolution imaging of $>$10\,keV X-rays, for which photoelectric absorption is negligible. Forthcoming hard X-ray telescopes such as {\em Astro-H} \citep{Takahashi:2010} and {\em NuSTAR} \citep{Harrison:2010} are expected to have such performance. Any $>$10\,keV X-ray emission centred on or peaking towards the centre of Core\,C may indicate the presence of penetrating high energy CRs, particularly if the energy spectrum of the X-rays hardens towards the core. If this effect results from suppression of CR diffusion as discussed in \S\,\ref{sec:Diffusion} the same suppression effect will apply to external electrons accelerated outside the core. Strong synchrotron radiative losses on the increasing magnetic field towards the core centre may further limit the X-ray component and/or alter the spectrum from these external electrons, to the effect that X-ray emission may appear somewhat edge- or limb-brightened in contrast to a more centrally peaked component from secondary electrons inside the core. For example the synchrotron cooling time $t_{\rm{sync}} \approx 1.5(B/ \rm{mG})^{-1.5}\,(\epsilon /\rm{keV})^{-0.5}$\,yr of electrons of energy $E$ responsible for X-ray photons of energy $\epsilon$ where $E \approx 62.5 \sqrt{ (\epsilon / \rm{keV}) / (B / \rm{mG})}$\,TeV, can be compared with the diffusion time $t_d \approx 16 \chi ^{-1} (R / \rm{1pc})^2 \sqrt{(B/ \rm{3\mu G})/(E/ \rm{GeV})}$\,yr into the core with radius, $R=0.62$\,pc. For hard X-ray photons $\epsilon \sim 10\,$keV and suppressed diffusion $\chi < 10 ^{-3}$ , we find that $t_{\rm{sync}}$ can become similar to or considerably less than $t_d$ as $B$ increases beyond $\gtrsim 80 \mu$G towards the centre of the core. For smaller values of $\chi$, $t_{\rm{sync}}$ becomes less than $t_d$ for even smaller values of $B$.

\section{Summary and Conclusion}
We used the Mopra 22\,m telescope to map in 7\,mm lines a region of gas towards the western rim of the gamma-ray-emitting SNR RX\,J1713.7$-$3946, encompassing the molecular cores A, B, and C (labeled by \citet{Moriguchi:2005}). Deep 7\,mm and 12\,mm pointed observations were also taken towards several of these regions (including Core\,D in the eastern rim) and we analysed archival 3\,mm data towards Core\,C. Our goals were to investigate the extent and properties of the dense gas components towards RX\,J1713.7$-$3946 and to complement previous CO studies of the low to moderately-dense gas \citep{Fukui:2003,Moriguchi:2005,Fukui:2008,Sano:2010}. The major conclusions from our observations are summarised as follows:

1) Detection of CS(J=1-0) emission towards three of the four cores sampled (Cores\,A, C, and D which have coincident infrared emission) confirming the presence of high density ($>10^4$\,cm$^{-3}$) gas. 

2) Detection of moderately broad CS(J=1-0) emission towards a position in between Cores\,A, B and C (labeled Point\,ABm) which is towards the X-ray outer shock. This broad gas may result from the SNR shock passing through, although no shock-tracing SiO emission was observed at this position.

3) The LTE mass for the brightest core, Core\,C, was estimated using three different molecular species, with results ranging from $\sim$40\,M$_{\odot}$ from CS observations to 80\,M$_{\odot}$ from NH$_3$ and N$_2$H$^+$ observations. The range of masses is most likely attributed to uncertainty in molecular abundances with respect to molecular hydrogen. Virial assumptions allowed molecular abundances with respect to molecular hydrogen of $\sim$5$\times$10$^{-8}$, $\sim$5$\times$10$^{-10}$ and $\sim$0.7 to 1$\times$10$^{-9}$ for NH$_3$, N$_2$H$^+$ and CS respectively to be calculated for Core\,C. Preliminary non-LTE \textsc{RADEX} modeling suggests a factor 2 lower density than our LTE results

4) Cores A and D were found to have LTE masses of 12 to 15\,M$_{\odot}$ and 30 to 120\,M$_{\odot}$, respectively, as traced by CS(J=1-0) emission, although these values are subject to core-radius uncertainties.

5) The inferred column density of our CS and N$_2$H$^+$ measurements towards Core\,C could lead to considerable photoelectric absorption ($\sim$95\%) of the $<5$\,keV X-ray emission. The small-scale keV X-ray features on the border of Core\,C therefore might not represent the complete X-ray emission towards this core. 

6) We investigated energy-dependent diffusion of TeV cosmic-ray protons into Core\,C using a simple model for the core based on average values of density and magnetic field. For the cases of suppressed diffusion coefficients with factors $\sim \chi=$10$^{-3}$ down to 10$^{-5}$, lower than the galactic average, considerable differences in the penetration depths between GeV and TeV CRs were found, with GeV CRs preferentially excluded. This effect could lead to a characteristic hardening of the TeV gamma-ray and hard X-ray emission spectrum peaking towards the core centre. Such features may be detectable and resolvable with future gamma-ray and X-ray telescopes and therefore offer a novel way to probe the level of accelerated CRs from RX\,J1713.7$-$3946 and the strength and structure of the magnetic field within the core. A more detailed diffusion model assuming a variable density and magnetic field is left for later work. We would also finally note that mapping in tracers of cosmic-ray ionisation such as DCO$^+$ and HCO$^+$ \citep{Montmerle:2010,Indriolo:2010} could a be highly valuable link between RX\,J1713.7$-$3946 and the observed gas and provide complementary information on the level of low energy MeV to GeV cosmic-rays impacting the molecular cores.



\section{Acknowledgments}
We would like to thank Stefano Gabici and Felix Aharonian for useful discussions about the diffusion of cosmic rays, and Malcolm Walmsley for his insightful comments. This work was supported by an Australian Research Council grant (DP0662810). The Mopra Telescope is part of the Australia Telescope and is funded by the Commonwealth of Australia for operation as a National Facility managed by the CSIRO. The University of New South Wales Mopra Spectrometer Digital Filter Bank used for these Mopra observations was provided with support from the Australian Research Council, together with the University of New South Wales, University of Sydney, Monash University and the CSIRO.

\appendix
\section{The Molecular Cores in Further Detail}\label{sec:AppA}
\subsection{C$^{34}$S and HC$_3$N emission}\label{sec:AppOtherMol}
\begin{figure}
\centering
\includegraphics[width=0.49\textwidth]{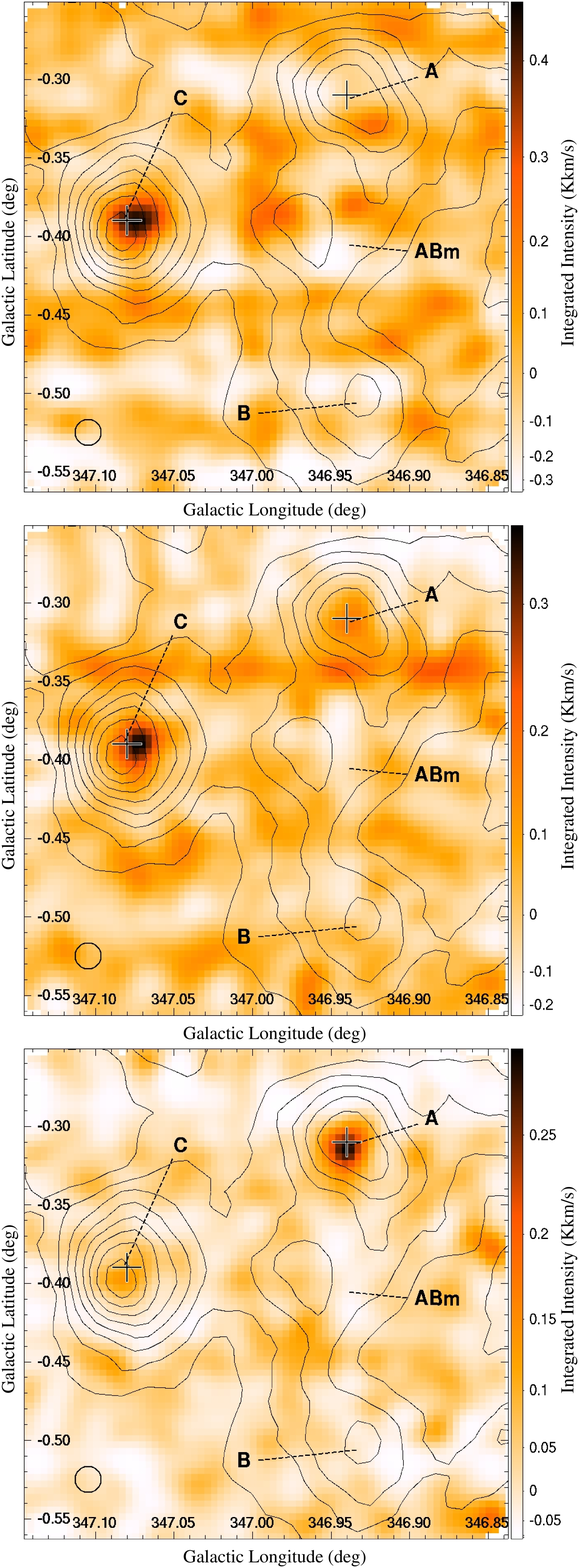}
\caption{\textbf{Top:} C$^{34}$S(J=1-0) v$_{\textrm{\tiny{LSR}}}=-$13.5~to~$-$9.5\,km\,s$^{-1}$. \textbf{Middle:} HC$_3$N(J=5-4) v$_{\textrm{\tiny{LSR}}}=-$13.5~to~$-$9.5\,km\,s$^{-1}$. \newline\textbf{Bottom:} HC$_3$N(J=5-4) v$_{\textrm{\tiny{LSR}}}=-$11.0~to~$-$9.0\,km\,s$^{-1}$.\label{fig:OtherSpecies}}
\end{figure}

C$^{34}$S(J=1-0) emission can be viewed towards Core\,C (Figure\,\ref{fig:OtherSpecies} top). This is further evidence of the existence of a dense molecular core and, with the CS(J=1-0) spectral line, allows the calculation of the optical depth and column density. 
HC$_3$N(J=5-4) emission from Cores A and C (Figure\,\ref{fig:OtherSpecies}, middle and bottom) are also indicative of dense cores.

\subsection{Scaling Factors}\label{sec:AppScalFact}
In this investigation, the Core\,C radius of 0.12\,pc, as calculated from CS(J=1-0) emission, has an uncertainty of $\sim$50\% and may not necessarily be representative of core sizes traced by NH$_3$(1,1) and N$_2$H$^+$(J=1-0) emission. This radius was also assumed for Core\,D. We therefore provide scaling-factors (Table~\ref{Table:ScalingFac}) normalised to a core radius of 0.12\,pc to allow parameter estimates derived from the three molecular species to be readily scaled to represent different source size assumptions. These were derived by simply calculating gas parameters for other core radius assumptions and dividing this by the result for a core radius of 0.12\,pc.

\begin{table}
\caption{Mass, $M$, density, n$_H$, and abundance, $\chi_{vir}$, scaling factors versus assumed Core\,C radius. Gas parameters derived from an assumed radius of 0.12\,pc can be readily converted to be for other assumed core radii by multiplying the value by the appropriate correction. \label{Table:ScalingFac}}
\begin{tabular}{lllllll}
\hline
Core Radius (pc)			&0.05 		&0.1	&0.15	&0.2	&0.25	&0.3\\
\hline
			$M$	 						&\multicolumn{6}{l}{Mass Scaling Factors}\\
			LTE 12\,mm		&5.7 		& 1.5 	& 0.68 	& 0.41 	& 0.28 	& 0.22\\
 			LTE 7\,mm		&5.3 		& 1.4 	& 0.72 	& 0.50 	& 0.41 	& 0.37\\
			LTE 3\,mm		&3.7 		& 1.2 	& 0.85 	& 0.76 	& 0.75 	& 0.74\\
			Virial			&0.42		&0.83	&1.25	&1.67	&2.08	&2.50\\
\hline
		 	$n_{H_2}$					&\multicolumn{6}{l}{Density Scaling Factors}\\
			LTE 12\,mm		&83 		& 2.7 	& 0.37 	& 0.09 	& 0.03 	& 0.01\\
 			LTE 7\,mm		&77 		& 2.6 	& 0.39 	& 0.11 	& 0.05 	& 0.03\\
			LTE 3\,mm		&53 		& 2.3 	& 0.46 	& 0.17 	& 0.09 	& 0.05\\
\hline			
			$\chi_{vir}$ &\multicolumn{6}{l}{Molecular Abundance Scaling Factors}\\
			12\,mm (NH$_3$)	&14 		& 1.8 	& 0.56 	& 0.25 	& 0.14 	& 0.09\\
			7\,mm	(CS) 	&13 		& 1.7 	& 0.59 	& 0.30 	& 0.20 	& 0.15\\
			3\,mm (N$_2$H$^+$)&8.9 		& 1.5 	& 0.69 	& 0.47 	& 0.36 	& 0.30\\
\hline
\end{tabular}
\end{table}

\subsection{Core\,C Radial Density Profile}\label{sec:AppRadProf}
\begin{figure}
\centering
\includegraphics[width=0.45\textwidth]{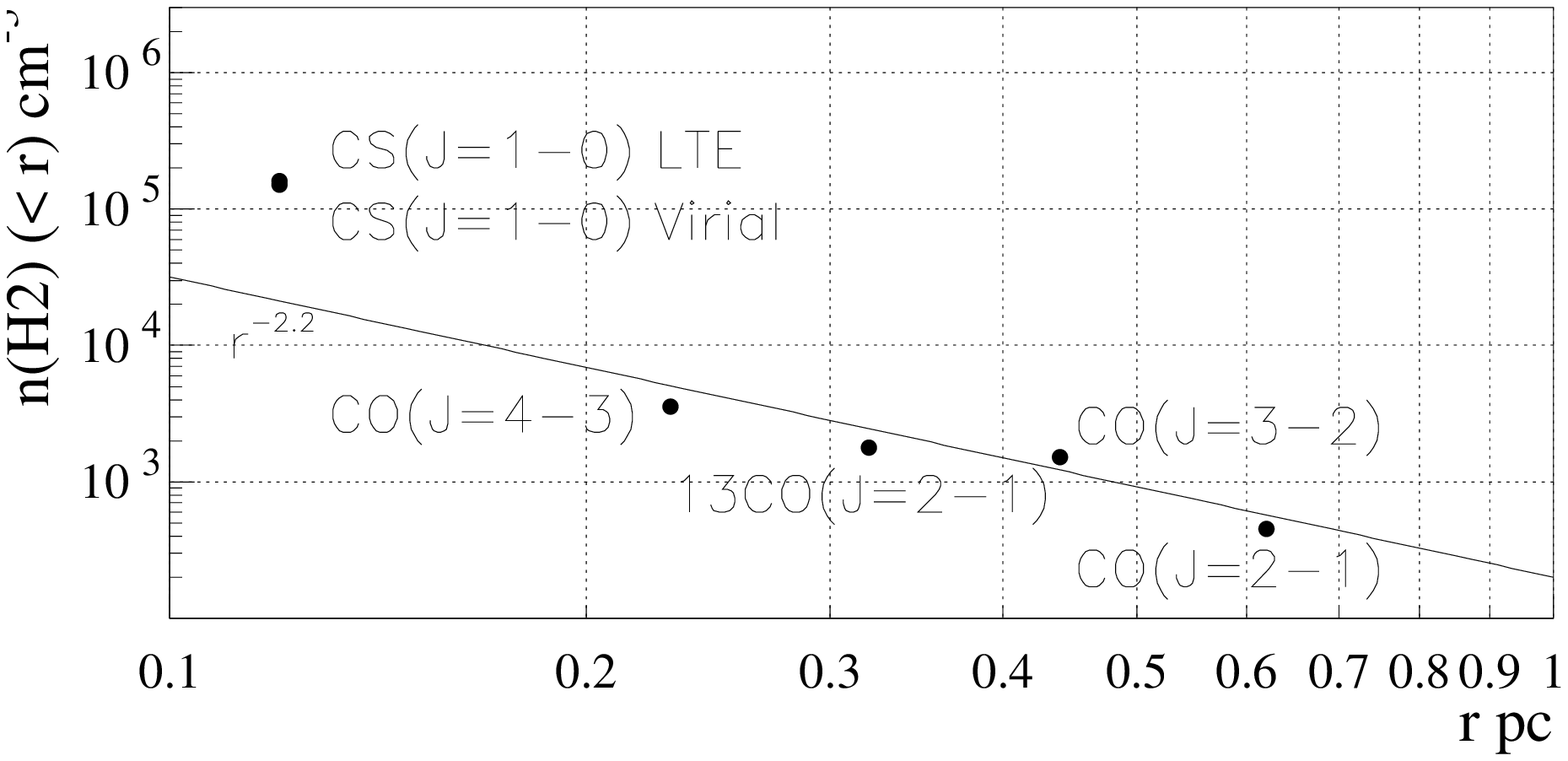}
\caption{Average density within radius, as traced by CO \citep{Sano:2010} and CS (this work). A $-$2.2 power-law function is also displayed for comparison (not fitted). Uncertainties are not displayed and are discussed in the text. \label{fig:radialdensity}}
\end{figure}

\citet{Sano:2010} used CO measurements to derive a power-law radial density profile of index $-$2.2$\pm$0.4 (Figure \ref{fig:radialdensity}). Extrapolating the model to $R<$0.12\,pc yields a number density value of $\sim$2$\times$10$^4$\,cm$^{-3}$. Our LTE results indicate a density of 2 to 3$\times$10$^5$\,cm$^{-3}$ and virial densities are 1 to 4$\times$10$^5$\,cm$^{-3}$, all an order of magnitude larger than that indicated by CO.

It follows that the Core\,C density profile may increase beyond the CO-derived power law at smaller radii, but this discrepancy could be due to abundance uncertainties discussed in \S\ref{sec:mass&abund}. This would not account for the discrepancy in virial-derived densities, which may simply be due to the virial assumption being invalid for Core\,C. If the apparent core radius varies between different molecular transitions, densities derived from different emission lines could be scaled independently (see Table\,\ref{Table:ScalingFac}), potentially changing estimates by an order of magnitude.

\subsection{CS Isotopologues}\label{sec:AppIsotop}
A key assumption in the optical depth calculations is that of the sulphur and carbon isotope ratio values, [CS]/[C$^{34}$S]$\sim$[$^{32}$S]/[$^{34}$S]$\sim$22.5 and [CS]/[$^{13}$CS]$\sim$[$^{12}$C]/[$^{13}$C]$\sim$75, respectively. These assumptions can be investigated by comparing the emission of $^{13}$CS and C$^{34}$S.  The $^{13}$CS(J=1-0)/C$^{34}$S(J=1-0) peak intensity ratio from Core\,C is 0.26$\pm$0.06, which would appear to be consistent with the adopted abundance ratios (22.5/75=0.3). On the other hand, the integrated intensity ratio is 0.16$\pm$0.05, somewhat inconsistent with our abundance assumptions. This is of concern as integrated intensity is more relevant than peak intensity when calculating molecular abundances. We reviewed $^{13}$CS(J=1-0)/C$^{34}$S(J=1-0) integrated intensity ratios in published literature towards a variety of cores.

\begin{figure}
\centering
\includegraphics[width=0.48\textwidth]{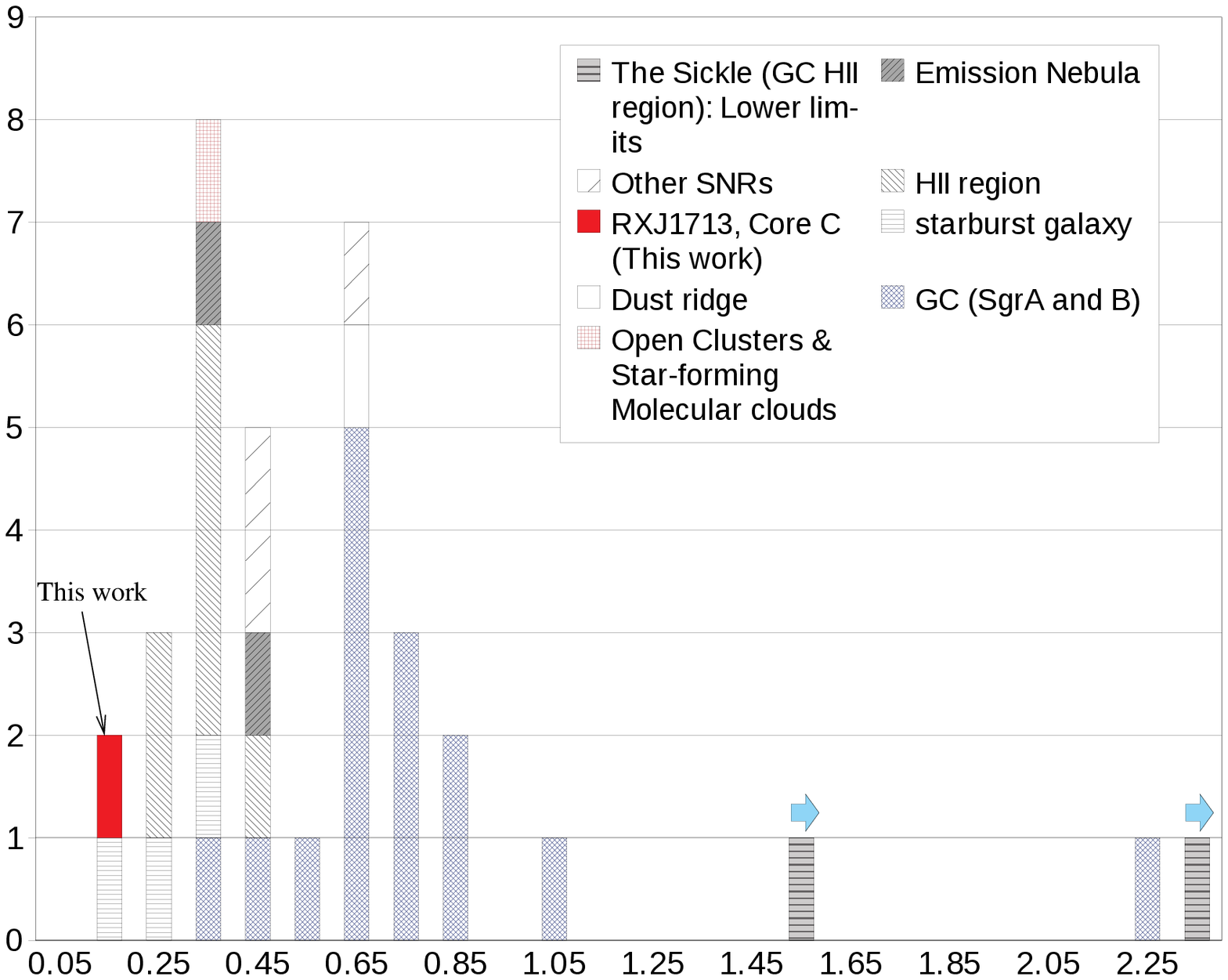}
\caption{$^{13}$CS/C$^{34}$S integrated intensity ratios for the J=1-0, J=2-1 and J=3-2 transitions. Data taken from \citet{Frerking:1980}, \citet{Martin:2008}, \citet{Martin:2005} and \citet{Nicholas7mm:2012}. Typical statistical error in data values is 25$\%$. \label{fig:13CSC34S}}
\end{figure}

Figure~\ref{fig:13CSC34S} presents a histogram of the ratio of integrated intensities of $^{13}$CS and C$^{34}$S emission from this sample. The graph suffers from low number statistics, but it can be seen that there is a large range of recorded [$^{13}$CS]/[C$^{34}$S] values, and although the Core\,C RX\,J1713.7$-$3946 data point is at the low end of the distribution, it represents only a $\sim$1$\sigma$ deviation from the average value of 0.55 (excluding the Sickle sources). Note-worthy in Figure \ref{fig:13CSC34S} are two galactic centre sources that recorded [$^{13}$CS]/[C$^{34}$S] values of over 4$\sigma$ from the mean. More data would be useful to compare the [$^{13}$CS]/[C$^{34}$S] ratio of Core\,C to that of other cores.

\section{The LTE Analyses is Detail}\label{sec:AppB}
\subsection{Ammonia Analysis}
\subsubsection{Optical Depth}
\begin{equation}
\frac{T_A^*(J,K, m)}{T_A^*(J,K,s)} = \frac{1-e^{-\tau(J,K,m)}}{1-e^{-\alpha\tau(J,K,m)}}
\end{equation}
\begin{equation}
\tau_{\rm J,K}^{\rm tot} = \frac{\tau(J,K,m)}{f(J,K)}
\end{equation}
\begin{equation}
\tau(2,2,m) = -{\rm ln} \left[ 1 - \frac{T_{A}^{*}(2,2,m)}{T_A^*(1,1,m)} \left( 1 - {\rm e}^{-\tau(1,1,m)} \right) \right]
\end{equation}

\subsubsection{Rotational \& Kinetic Temperature}

\begin{equation}
T_{rot} = -41 \left[ \ln \left( \frac{9}{20}\frac{\tau_{2,2}^{\rm tot}}{\tau_{1,1}^{\rm tot}} \right) \right]^{-1}\hspace{3mm}[\rm{K}]
\end{equation}

\begin{equation}
T_{kin} = \frac{T_{rot}}{1-\frac{T_{rot}}{42}~{\rm ln} \left[ 1+ 1.1~\textrm{exp}\left(-\frac{16}{T_{rot}}\right) \right]}\hspace{3mm}[\rm{K}]
\end{equation}

\subsubsection{Column Density}
\begin{equation}
N_{1,1} = \frac{8 k \pi \nu^{2}}{A_{u} h c^{3}}\left(\frac{\Delta\Omega_a}{\Delta\Omega_s}\right)\left(\frac{\tau_{1,1}^{\rm tot}}{1-e^{-\tau_{1,1}^{\rm rot}}}\right)\hspace{2mm}\int T_{mb}~dv\hspace{3mm}[\rm{cm}^{-2}]
\end{equation}
\begin{equation}
N_{\textrm{\tiny{NH}}_{3}}=\frac{N_{1,1}}{g_{1,1}}e^{E_{1,1}/kT}Q(T)\hspace{3 mm}[\rm{cm}^{-2}]
\end{equation}
\begin{equation}
Q(T) = \Sigma_{i} g_{i}e^{-E_{i}/kT}
\end{equation}

\subsection{N$2$H$^+$ Analysis}
\subsubsection{Optical Depth}
\begin{equation}
\frac{T_{main}}{T_{hf}}=\frac{1-e^{-\tau _{main}}}{1-e^{-\alpha\tau _{main}}}
\end{equation}
\begin{equation}
\tau_{tot} = \frac{\tau _{main}}{f}
\end{equation}
\subsubsection{Excitation Temperature}
\begin{equation}
T_{ex}=\frac{T_{mb}}{\left(\frac{\Delta\Omega_a}{\Delta\Omega_s}\right)(1-e^{-\tau})} + \left(\frac{\Delta\Omega_s}{\Delta\Omega_a}\right)T_b
\end{equation}
\subsubsection{Column Density}
\begin{equation}
N_{N_2H^+}=\frac{3\sqrt{\pi}h\epsilon_0}{2\pi^2\sqrt{\ln{2}}\mu_{N_2H^+}^2}
\frac{e^{\frac{h\nu}{kT_{ex}}}}{e^{\frac{h\nu}{kT_{ex}}}-1}
\left( \frac{kT_{ex}}{h\nu} + \frac{1}{6}\right) 
\Delta V_{\rm{fwhm}}\tau
\end{equation} 

\subsection{CS}
\subsubsection{Optical Depth}
\begin{equation}
\frac{T _{CS}}{T _{CS~iso.}} = \frac{1-e^{-\tau _{CS}}}{1-e^{-\alpha \tau _{CS}}}
\end{equation} where $\alpha$ is the ratio of the rarer CS isotopologue abundance and CS abundance.
\subsubsection{Column density}
\begin{equation}
N_{CS(J=u)} = \frac{8k\pi \nu_{ul} ^2}{A_{ul}hc^3} \left(\frac{\Delta\Omega_a}{\Delta\Omega_s}\right)  \int T_{mb} dv \frac{\tau}{1-e^{-\tau}}
\end{equation} 
\begin{equation}
N_{CS} =  \frac{N_{CS(J=1)}}{g_{CS(J=1)}}e^{E_{CS(J=1)}/kT}Q(T_{rot})\hspace{3 mm}[\rm{cm}^{-2}]
\end{equation}

\subsubsection{Rotational Temperature}
\begin{equation}
T_{rot} = -4.7 \left[ \ln \left( \frac{3}{5}\frac{N_{CS(J=2)}}{N_{CS(J=1)}} \right) \right]^{-1}\hspace{3mm}[\rm{K}]
\end{equation}

\subsection{LTE mass, Density and Virial Abundance}
\begin{equation}
M=\frac{2\pi r^2 m_HN_{Mol}}{\chi}
\end{equation}
\begin{equation}
n_{H_2}=\frac{N_{Mol}}{\chi r}
\end{equation}
\begin{equation}
\chi _{vir} = \frac{2\pi r^2 m_HN_{Mol}}{M_{vir}}
\end{equation}





\end{document}